\documentclass[aps,prx,twocolumn,superscriptaddress]{revtex4-1}
\pdfoutput=1
\usepackage{amsmath}
\usepackage{amssymb}
\usepackage{graphicx}
\usepackage{enumerate}
\usepackage{caption}
\usepackage{mathrsfs}
\usepackage{braket}
\usepackage{bm}
\usepackage{comment}
\usepackage[sort&compress]{natbib}
\usepackage{caption}
\usepackage{url}
\usepackage[lofdepth,lotdepth]{subfig}
\usepackage[colorlinks=true]{hyperref}
\usepackage{multirow}

\def\cu2{$\text{Cu}_2\text{OSeO}_3$}

\allowdisplaybreaks
\usepackage[table]{xcolor}
\usepackage{svg}
\begin{document}

\title{Low energy magnons in the chiral ferrimagnet $\text{Cu}_2\text{OSeO}_3$: a coarse-grained approach}

\author{Yi Luo}
\email{yluo13@jhu.edu}
\affiliation{
	Institute for Quantum Matter and Department of Physics and Astronomy, Johns Hopkins University, Baltimore, Maryland 21218, USA}
\author{G. G. Marcus}
\affiliation{
	Institute for Quantum Matter and Department of Physics and Astronomy, Johns Hopkins University, Baltimore, Maryland 21218, USA}
\author{B. A. Trump}
\affiliation{
NIST Center for Neutron Research, National Institute of Standards and Technology,
Gaithersburg, Maryland 20899-6102, USA}
\author{J. Kindervater}
\affiliation{
	Institute for Quantum Matter and Department of Physics and Astronomy, Johns Hopkins University, Baltimore, Maryland 21218, USA}
\author{M. B. Stone}
\affiliation{Neutron Scattering Division, Oak Ridge National Laboratory, Oak Ridge, Tennessee 37831, USA}
\author{J. A. Rodriguez-Rivera}
\affiliation{
	NIST Center for Neutron Research, National Institute of Standards and Technology,
	Gaithersburg, Maryland 20899-6102, USA}
\affiliation{
Department of Materials Science and Engineering, University of Maryland, College Park, MD, 20742, USA}
\author{Yiming Qiu}
\affiliation{
	NIST Center for Neutron Research, National Institute of Standards and Technology,
	Gaithersburg, Maryland 20899-6102, USA}
\author{T. M. McQueen}
\affiliation{
	Institute for Quantum Matter and Department of Physics and Astronomy, Johns Hopkins University, Baltimore, Maryland 21218, USA}
\affiliation{Department of Materials Science and Engineering, Johns Hopkins University, Baltimore, Maryland 21218, USA}
\author{O. Tchernyshyov}
\affiliation{
	Institute for Quantum Matter and Department of Physics and Astronomy, Johns Hopkins University, Baltimore, Maryland 21218, USA}
\author{C. Broholm}
\email{broholm@jhu.edu}
\affiliation{
	Institute for Quantum Matter and Department of Physics and Astronomy, Johns Hopkins University, Baltimore, Maryland 21218, USA}
\affiliation{Department of Materials Science and Engineering, Johns Hopkins University, Baltimore, Maryland 21218, USA}
\affiliation{
	NIST Center for Neutron Research, National Institute of Standards and Technology,
	Gaithersburg, Maryland 20899-6102, USA}

	\begin{abstract}
	We report a comprehensive neutron scattering study of low energy magnetic excitations in the breathing pyrochlore helimagnetic $\text{Cu}_2\text{OSeO}_3$. Fully documenting  the four lowest energy magnetic modes that leave the ferrimagnetic configuration of the ``strong tetrahedra'' intact ($|\hbar\omega|<13$~meV), we find gapless quadratic dispersion at the $\Gamma$ point for energies above 0.2 meV, two doublets separated by 1.6(2) meV at the $R$ point, and a bounded continuum at the $X$ point. Our constrained rigid spin cluster model relates these features to Dzyaloshinskii-Moriya (DM) interactions and the incommensurate helical ground state. Combining conventional spin wave theory with a spin cluster form-factor accurately reproduces the measured equal time structure factor through multiple Brillouin zones. An effective spin Hamiltonian describing the complex anisotropic inter-cluster interactions is obtained. 	
\end{abstract}

\maketitle

\section{Introduction}
	Chiral magnets have attracted a great deal of attention for a long time \cite{dzyaloshinskii1964,bak1980theory,muhlbauer2009skyrmion}. The absence of inversion symmetry in the atomic lattice gives rise to twists of magnetization $\mathbf M(\mathbf r)$ in magnetically ordered states, which range from simple helices to intricate periodic lattices of skyrmions and magnetic hedgehogs. The microscopic mechanism responsible for the twisting of magnetization is the spin-orbit coupling manifesting itself in magnetic insulators as the Dzyaloshinskii-Moriya (DM) interaction of the form $\mathbf M \cdot (\nabla \times \mathbf M)$ in the continuum approximation \cite{dzyaloshinskii1964}. On the atomistic level, the DM interaction is represented by the pairwise spin interaction $\mathbf D_{ij} \cdot (\mathbf S_i \times \mathbf S_j)$, where $\mathbf D_{ij}$ is a vector specific to the bond connecting spins $\mathbf S_i$ and $\mathbf S_j$ \cite{moriya1960}. Determination of spin interactions in chiral magnets is very important for the understanding of their magnetic states. 
	
	We present an experimental study of the chiral magnet \cu2 by means of inelastic neutron scattering. This compound has a cubic lattice symmetry without an inversion center (space group $P2_13$) \cite{belesi2011magnetic} and exhibits paramagnetic, helical, conical, and skyrmion-crystal phases as a function of temperature and applied magnetic field \cite{adams2012long,white2012electric,seki2012formation,reim2017impact,makino2017thermal,bannenberg2017reorientations,white2018electric,qian2018new,chacon2018observation}. The structural unit cell has 16 magnetic Cu$^{2+}$ spin-1/2 ions which makes a microscopic description at the level of individual spins rather complex and impractical. Romhanyi \emph{et al.} \cite{romhanyi2014entangled,ozerov2014establishing,portnichenko2016magnon,tucker2016spin} introduced a microscopic model with Heisenberg exchange interactions of five different strengths: $J_s^{\text{AF}},J_s^{FM},J_w^{\text{AF}},J_w^{\text{FM}},J^{\text{AF}}_{\text{o.o}}$(FM and AF represent ferromagnetic and antiferromagnetic interactions, respectively), shown in Fig.~\ref{Fig1}(a). As will be shown below, this model nontheless misses significant features of the low energy magnon spectrum. While these problems might be remedied by the addition of DM interactions, a further increase in complexity would be undesirable. 
	
	Fortunately, magnetic interactions in \cu2 exhibit a hierarchy of energy scales \cite{romhanyi2014entangled,janson2014quantum,grigoriev2019spin}, which allow for an efficient modeling at a coarse-grained level, wherein quartets of strongly interacting spins are treated as effective spins with weaker interactions between them. Hints of this hierarchy can be seen in the inelastic neutron spectrum shown in Fig.~\ref{Fig1}(b). It reveals four strongly dispersing magnon bands at low energies (0-12 meV) separated by a large gap from high-energy magnon bands with a relatively weak dispersion (25-33 meV). The low-energy branches are spin waves where spins within each strongly coupled tetrahedron precess in phase with each other and can be described by a single effective spin within a coarse-grained model [Fig.~\ref{Fig1}(c,d)], while the high-energy magnons are associated with the intra-cluster interactions. To bring out the interactions that are relevant for the complex phase diagram and ordered structures, we focus on the low energy inter-cluster magnons in our study.  The coarse-grained picture we adopt enables us to identify and refine the magnitude of the anisotropic interaction terms relevant to the helical and skyrmionic spin textures in \cu2. We show these terms can be gleaned from specific features in high resolution neutron scattering spectra at energies well beyond the collective energy scales of the mesoscopic phases. We also show how to define the relevant low-energy degrees of freedom for a complex magnetic material with a hierarchy of energy scales and provide a simple expression for the corresponding inelastic scattering cross section in terms of a cluster form factor.  \par
	
	The paper is organized as follows: In Sec.~\ref{Exp} we present our detailed inelastic magnetic neutron scattering data for \cu2 with a focus on the new features that they reveal in the low energy regime. These features will then be related to DM interactions and the associated incommensurate ground state through the simplified coarse-grained model introduced in Sec. \ref{theo}. In Sec. \ref{Num} we numerically calculate the structure factors after deriving the effective form factor (details in Appendix~\ref{Formderive}), and determine the set of interaction parameters by a pixel to pixel data fit. The resulting best-fit parameters are listed in Table \ref{Pa}, bolstered by a detailed discussion of the reliability of the fit and the corresponding error bars in Appendix \ref{Reliability}. The power of the effective model and its limitations are identified and discussed in Sec.~\ref{Dis} before concluding in Sec.~\ref{Concl}.\par
	
	Throughout this paper, we use the same lattice structure conventions of Janson et al.~\cite{janson2014quantum}, where the coordinates of 16 Cu ions within the unit cell of a right-handed enantiomer are listed. These are reproduced in Table \ref{Posit} of Appendix.~\ref{LWST}.
\begin{figure*}[htbp!]  
	\includegraphics[width=0.8\textwidth ]{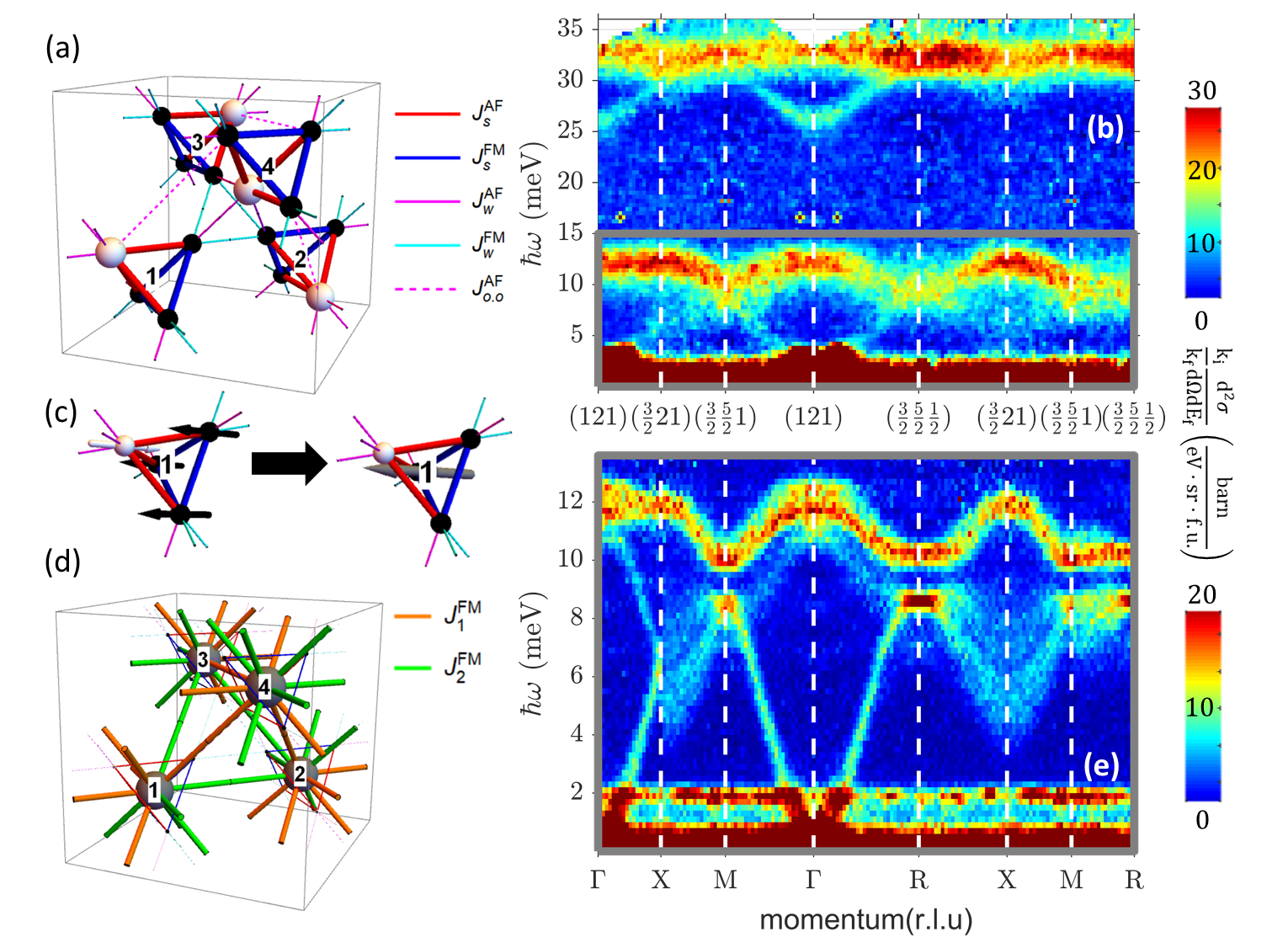}
	\caption{(a): Structure of the right-handed enantiomer of cubic \cu2. ($a=8.911~\text{\AA}$ space group $P2_13$ \cite{bos2008magnetoelectric,belesi2011magnetic}). Each unit cell contains 16 $\text{Cu}^{2+}$ ions. The two distinct $\text{Cu}^{2+}$ sites are labeled by Cu-1 (white) and Cu-2 (black), respectively. $J_s^{\text{AF}}$ (blue, thick) and $J_s^{\text{FM}}$ (red, thick) are the dominant magnetic interactions. (b) The measured inelastic magnetic neutron scattering  cross section acquired with incident neutron energy $E_i=60$ meV at $T=4$~K. The 4D data set is displayed as slices along a trajectory in momentum space connecting the high symmetry points $\Gamma(h,k,l)$; $X(h,k,l+\frac{1}{2})$; $M(h,k+\frac{1}{2},l+\frac{1}{2})$; and $R(h+\frac{1}{2},k+\frac{1}{2},l+\frac{1}{2})$. Here, $h,k$, and $l$ are integers. The integration range of perpendicular $\mathbf{Q}$ direction is 0.1 $\text{\AA}^{-1}$. (c) Each strong tetrahedron is composed of one Cu-1 and three Cu-2 sites, with AF interactions between Cu-1 and Cu-2 sites, and FM interactions between Cu-2 sites. This results in an effective spin-1 cluster with a Cu-1 spin antiparallel with three parallel Cu-2 spin. (d) The effective spins occupy a distorted FCC lattice with effective ferromagnetic inter-cluster interactions. We define the sites connected by the bonds  $J_1^{\text{FM}}$ and $J_2^{\text{FM}}$ to be nn and nnn, respectively. (e)  The measured inelastic magnetic neutron scattering cross section acquired with $E_i=20$ meV, focusing on the energy range indicated by the gray box in (b). (e) shows the average intensity along the indicated trajectories in the Brillouin zones centered at (021),(111),(121) and (122) averaging over $\pm 0.05~\mathrm{\AA^{-1}}$ in perpendicular $\bf Q$-directions. For (111) only data with energy transfers below 10.5 meV is taken into the average since data with higher energy transfer is not covered well due to kinematic limitations. Four magnon modes are generally observed corresponding to four clusters per unit cell. Additional modes can result from down-folding due to the incommensurate helimagnetic ground state and domain averaging. The intensity band at 2 meV arises from a spurious process unrelated to \cu2.}
	\label{Fig1}
\end{figure*}

\section{Inelastic neutron scattering}\label{Exp}
 Single crystals of \cu2 were grown by chemical vapor transport. Approximately 50 crystals were co-aligned on an aluminum holder for a total sample mass $m\approx5.1$ g and full width at half maximum (FWHM) mosaic $\approx 0.5^{\circ}$. No provision was made to check individual crystal chirality or orientation apart from aligning the four fold axes so the overall symmetry of the mosaic has approximate cubic symmetry. Time-of-flight inelastic neutron scattering data were acquired on the SEQUOIA instrument at the Spallation Neutron Source. Incoming neutron energies of $E_i=60$ meV and 20 meV were used with the high flux chopper operating at 240 Hz and the high resolution chopper operating at 180 Hz, respectively. The corresponding FWHM elastic energy resolution was 3 meV and 0.5 meV, respectively. The data were acquired at $T=4$~K which is far below the critical temperature $T_c=58$~K. The sample was cooled using a closed-cycle refrigerator, and rotated through $180^\circ$ in $0.5^\circ$ steps about the $(h\bar{h}0)$ axis. These same spectrometer settings were used to measure
 vanadium incoherent scattering for absolute normalization of
 the differential scattering cross section. The total beam time accumulated was 0.0655 Ah for  $E_i=60$ meV and 0.0673 Ah for $E_i=20$~meV. The data were analyzed in Mantid \cite{arnold2014mantid} where background contributions were masked and subsequently symmetrized in the m$\bar{3}$m Laue class using Horace \cite{ewings2016horace}.\par
 	\begin{figure*}[!htbp] 
 	\includegraphics[width=\textwidth]{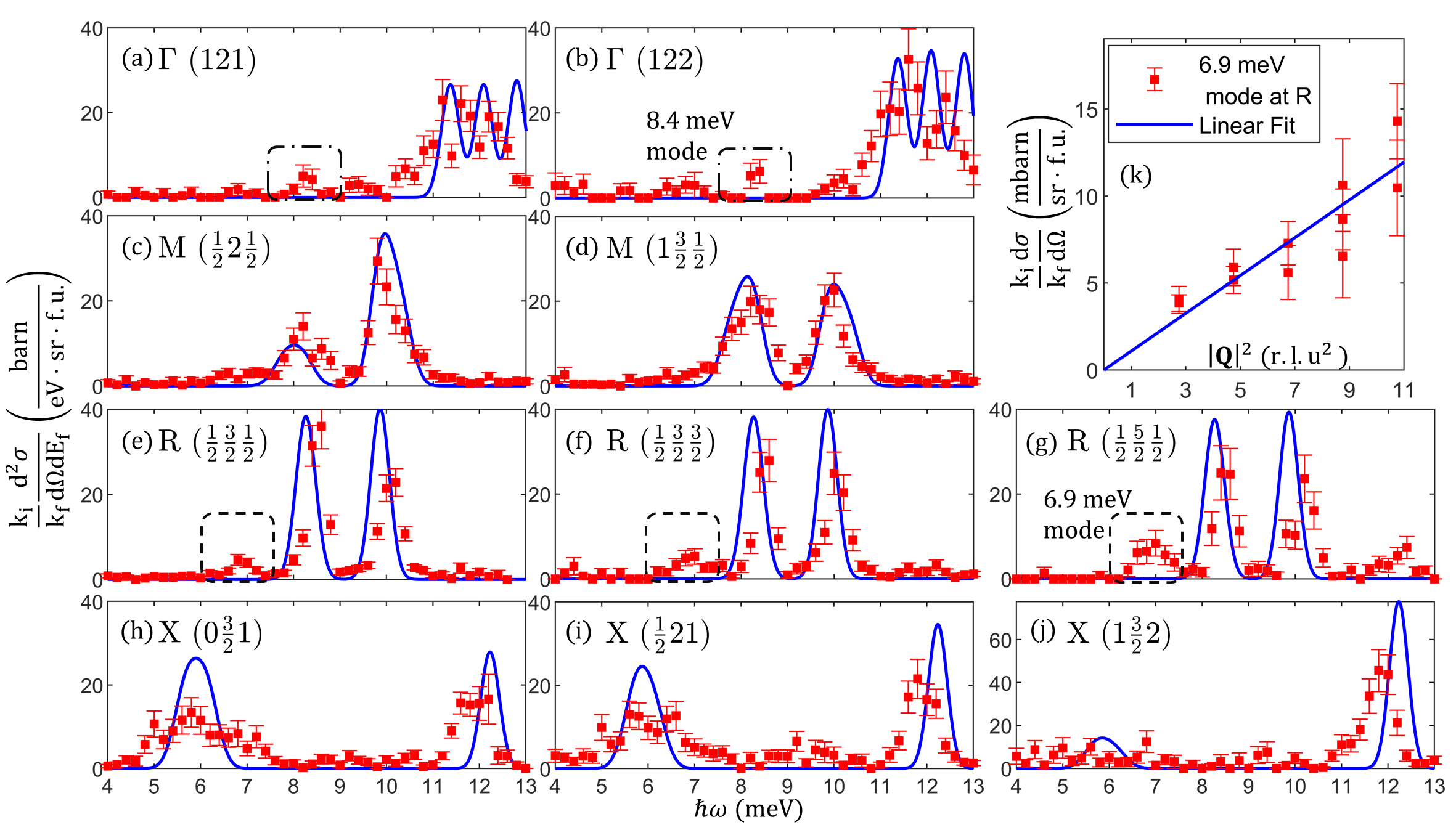}
 	\caption{(a-j) Inelastic magnetic neutron scattering spectra for $\rm Cu_2OSeO_3$ acquired for T=4 K at high symmetry points in the Brillouin zone. Red symbols show  neutron intensity data averaged over $(0.084~\text{\AA}^{-1})^3\times(0.2~\text{meV})$ in the 4D $\mathbf{Q}-\hbar\omega$ space. The blue line shows the result of a highly constrained calculation of the scattering cross section associated with spin waves described by the effective spin-1 model with the optimized exchange parameters listed in Table~\ref{Pa}. The FWHM of the peaks (blue) was determined from instrument energy resolution and a phenomenological relaxation rate $\tilde{\Gamma}=0.19$~meV to characterize on average the extra physical broadening throughout the Brillouin zone (see Sec.~\ref{Num} and Appendix~\ref{reso}). Note the excess broadening of the lower mode at the $X$ point (h-j), which we ascribe to two magnon decay processes that are kinematically accessible here and effectively destroy the $X$ point single magnon (Fig.~\ref{Fig2}). As discussed in Sec.~\ref{theo}, we expect two two-fold degenerate modes at  $R$. In the measured cross section at high momentum, a third mode at 6.9 meV can also be observed. The intensity of this mode averaged over $(0.084~\text{\AA}^{-1})^3$ and integrated over [6,7.8] meV is plotted versus $|\mathbf{Q}|^2$ in (k). The linear fit  indicates this mode is a phonon. The 8.4 meV modes marked in (a,b) were discussed in Ref.\cite{laurita2017low}. Error bars in all figures represent one standard deviation.}
 	\label{Fig3}
 \end{figure*}
The $E_i=60$ meV inelastic neutron scattering cross section in Fig.~\ref{Fig1}(b) shows a large ($\approx 13$ meV) energy gap separating the four lowest branches from  higher energy modes.
The $E_i=20$ meV data are displayed as a false-color image in Fig.~\ref{Fig1}(e) and as energy cuts at representative high symmetry points $R(\frac{1}{2}, \frac{5}{2},\frac{1}{2})$, $X(1,2,\frac{1}{2})$, $M(\frac{1}{2},2,\frac{1}{2})$, and $\Gamma(1,2,2)$ in Fig.~\ref{Fig3}. The high symmetry points are defined as: $\Gamma(h,k,l)$; $X(h,k,l+\frac{1}{2})$; $M(h,k+\frac{1}{2},l+\frac{1}{2})$; and $R(h+\frac{1}{2},k+\frac{1}{2},l+\frac{1}{2})$ with $h,k$, and $l$ integers. While broadly consistent with the prior work \cite{romhanyi2014entangled}, our high-resolution data reveal important new features: (1) A splitting at the $R$ point $\Delta_R=1.6(2)$ meV between the two modes with dominant intensity(previously reported by \textcite{tucker2016spin}), whereas the Heisenberg model of \textcite{romhanyi2014entangled} implies four-fold degeneracy. A third mode between 6 meV and 8 meV can also be observed at  $R$ points for high momentum transfer. Consistent with Ref.\cite{tucker2016spin}, we identify this mode as a phonon (Fig.~\ref{Fig3}(k)) based on the $|\textbf{Q}|^2$ dependence of the integrated intensity\cite{Zaliznyak_2014}  (2) Near the $X$ point there is a dramatic broadening of the lower branch (between 4 and 8 meV in Fig~\ref{Fig1}(e)), where the Heisenberg model \cite{romhanyi2014entangled} calls for two-fold degeneracy. (3) The optical modes at the $\Gamma$ point at 11.6 meV, which in the Heisenberg model is triply degenerate, is split into three modes with splitting $\Delta_{\Gamma}^{o}=0.7(3)$ meV, see Sec.\ref{alge}. In the following we will show that these features directly reflect symmetry-allowed DM interactions and the associated incommensurate nature of the ground state. \par

As apparent in Fig.~\ref{Fig1}(e), the low energy parts ($<2$ meV) of the inelastic magnetic scattering at $\Gamma$ points overlap with the tails of elastic coherent and incoherent nuclear and magnetic scattering as a result of the finite energy resolution of the measurements. To resolve magnetic scattering in this low energy regime, we used the MACS instrument\cite{Rodriguez_2008} at the NIST Center for Neutron Research in a separate experiment on the same sample. The final energy was fixed at $E_f=2.4$ meV resulting in a FWHM elastic energy resolution 0.08 meV. The data were acquired at $T=1.6$ K.  We were able to resolve magnon dispersion with energy transfers from $\hbar\omega=0.2$ meV to 1.2 meV. The data were processed using the software DAVE\cite{Azuah_2009} and folded assuming cubic symmetry.\par

\begin{figure}[!htbp] 
	\includegraphics[width=\columnwidth]{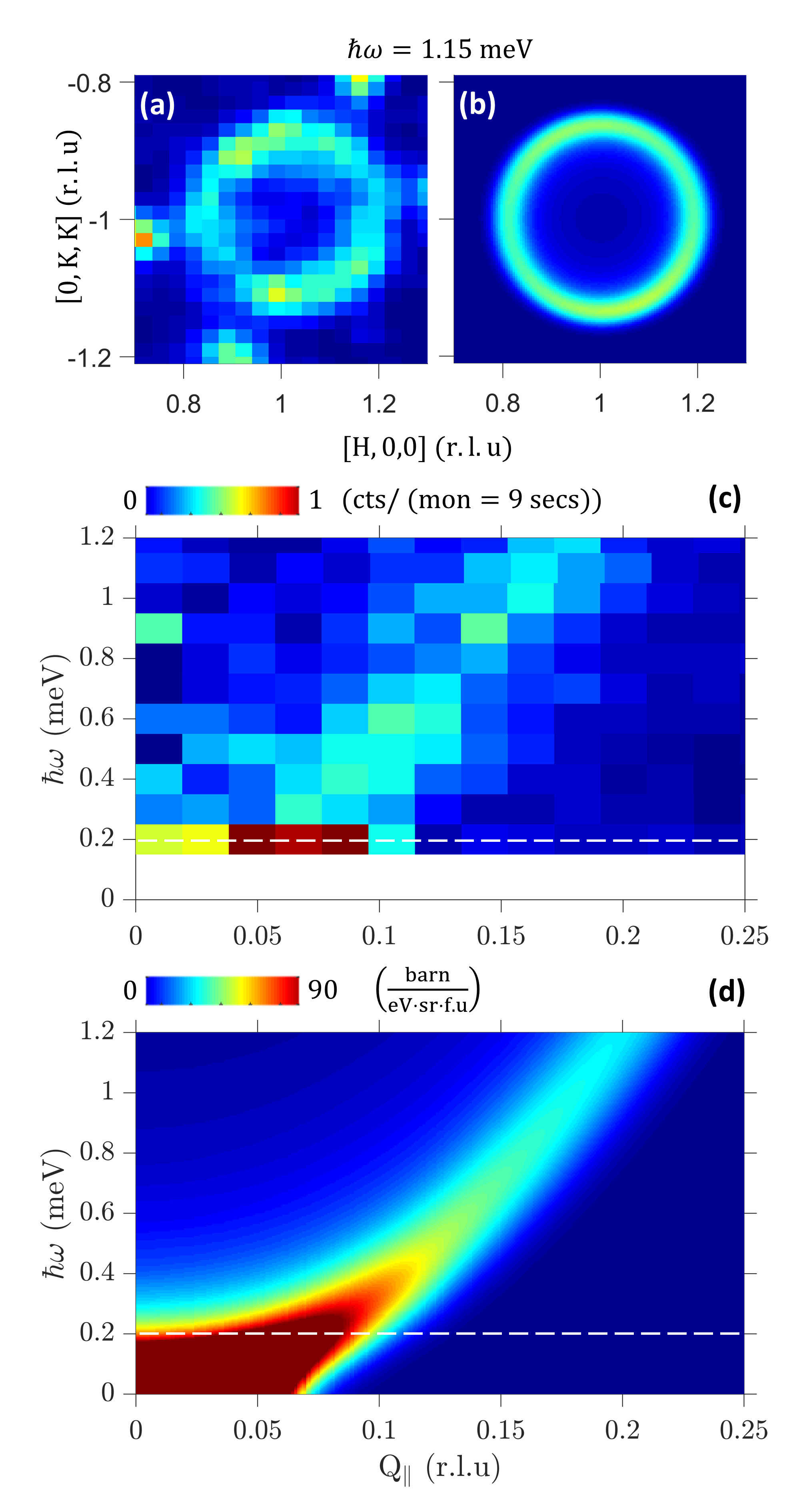}
	\caption{(a) Constant $\hbar\omega=1.15(15)$ meV slice through MACS data near the ${\bf Q}_0=(1\bar{1}\bar{1})$ zone center. The spinwave signal forms a circle, which indicates isotropic dispersion. (b) Spinwave model calculation using the parameters in Table.~\ref{Pa} and numerically convoluting with the instrumental resolution described in Appendix~\ref{MACSapp}. (c) $Q_\parallel-\omega$ intensity map of MACS data following azimuthal averaging around  ${\bf Q}_0$. Due to the azimuthal averaging, the errorbars of the pixels are inversely proportional to $Q_{\parallel}$; The pixels near $Q_{\parallel}=0$ (for example, bright pixels at $\hbar\omega=0.4,0.6,0.9$ meV) have significantly larger errorbars compared to the pixels of finite $Q_{\parallel}$ and are thus less reliable. (d) Calculated $Q_\parallel-\omega$ intensity map using parameters in Table.~\ref{Pa} and the same azimuthal averaging as for the experimental data. Data in (a,c) shares the same color scal and was not independently normalized. Calculation results in (b,d) share the same normalized color scale. Dashed lines in (c,d) marks the lowest accessible energy transfer (0.2 meV) in the MACS experiment.}
	\label{MACS1}
\end{figure}

A fixed $\hbar\omega=1.15$ meV slice of MACS data near the $\Gamma(1,\bar{1},\bar{1})$ zone center is shown in Fig.~\ref{MACS1}(a). Within experimental accuracy, the  dispersion is  isotropic. Notice the four point-like signals outside the rings in Fig.~\ref{MACS1}(a). These are remnants of Bragg diffraction of 2.4 meV neutrons diffusely scattered from the monochromator that were partially subtracted as described in Appendix~\ref{spur}. We approximate the dispersion as $E(q)=Dq^2+\Delta_{\Gamma}$, where $q$ is the distance from  the $\Gamma$ point, $D$ is the spinwave stiffness and $\Delta_{\Gamma}$ is a possible anisotropy gap. Taking into account the coarse out of plane Q-resolution of MACS and its energy resolution as described in Appendix~\ref{MRcal}, a pixel-to-pixel fit to the data yields $D=67(8)~\text{meV}~ \text{\AA}^2$, which is slightly larger than the previous neutron report\cite{portnichenko2016magnon} and the overall model parameters in Table~\ref{Pa}, which fit the SEQUOIA data of higher energy transfers and correspond to $D=58(2)~\text{meV}~ \text{\AA}^2$  where the latter range indicates the orientational anisotropy. The data place an upper bound of 0.1 meV on $\Delta_{\Gamma}$, which is consistent with other experiments\cite{kobets2010microwave,prasai2017ballistic}. Fig.~\ref{MACS1}(c,d) compare the angular average neutron scattering intensity data to the resolution smeared intensity distribution anticipated for the best-fit coarse grained model indicated in Table~\ref{Pa}. Here the effects of momentum and energy resolution were taken into account as described in Appendix~\ref{MACSapp} where we also discuss evidence for the incommensurate ground state in the form of a physical momentum space broadening of low energy modes.\par

\section{Spinwave model}\label{theo}
\begin{figure}[!htbp] 
	\includegraphics[width=\columnwidth]{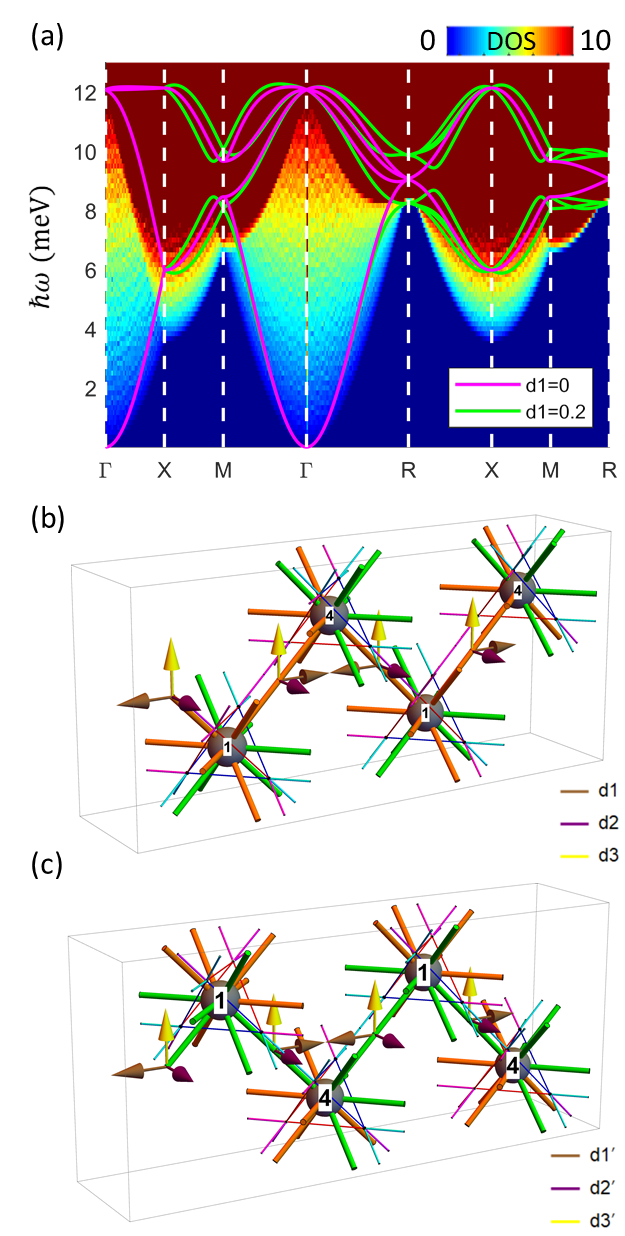}
	\caption{(a) Magnon dispersion calculated for $\mathcal{H}_J$ with $J_1=-0.6$ meV, $J_2=-0.9$ meV (magenta), and for $\mathcal{H}_{\text{tot}}\equiv\mathcal{H}_J+\mathcal{H}_D$ with $d_1=-d_1'=0.2$ meV and all other DM components zero (green). The general features of the $E_i=20$ meV inelastic neutron data (Fig.~\ref{Fig1} and Fig.~\ref{Fig3}) are reproduced. The DM interactions lift the $R$ point degeneracy as observed in the experimental data (Fig.~\ref{Fig3}(e-g)). The colored background shows the density of states (DOS) of the 2-magnon continuum for each momentum along high symmetry direction. The unit for the DOS is $1/(\text{\AA}^{-1}~\text{meV})$ per unit cell. (b) Local coordinate system defining DM interactions for nn and nnn effective spins. Only the DM interaction for a single nn pair $\mathbf{D}_{14}$ is shown. For nnn (c) a similar set of $(d_1',d_2',d_3')$ projections can be defined. There are no symmetry constraints on $\mathbf{D}$ or $\mathbf{D}^\prime$. Only the components $d_1$ and $d_1'$ contribute to splitting at the $R$ point.}
	\label{Fig2}
\end{figure}

Without compromising accuracy, great simplification in modeling the low-energy spin dynamics of \cu2 can be achieved by treating each strong tetrahedron as a rigid cluster with an effective spin $S=1$. The corresponding coarse-grained lattice shown in Fig.~\ref{Fig1}(d) is a distorted FCC lattice with the same space group $P2_13$ as the original lattice. There are two different types of ferromagnetic interaction between the effective spins. As shown in Fig.~\ref{Fig1}(a,d), we define the bond arising from $J_w^{\text{AF}}$ and $J_{\text{o.o}}^{\text{AF}}$ to be $J_1$ (nearest neighbor, nn). The interaction arising from $J_w^{\text{FM}}$ is denoted $J_2$ (next nearest neighbor, nnn). The Hamiltonian for the effective model reads
\begin{equation}
\mathcal{H}_J = \sum_{\langle ij \rangle}J_1\mathbf{S}_i\cdot\mathbf{S}_j 
	+ \sum_{\langle \langle ij \rangle \rangle}J_2 \mathbf{S}_i\cdot\mathbf{S}_j,
\end{equation}
where $\langle ij \rangle$ and $\langle \langle ij \rangle \rangle$ denote pairs of first and second neighbors, respectively. 
We then use the standard Holstein-Primakoff (HP) substitution for collinear structures and expand to order of $1/S$ before setting $S=1$. The dispersion relation for the resulting quadratic magnon hopping model (Fig.~\ref{Fig2}(a)) is broadly consistent with the inelastic neutron scattering data in Fig.~\ref{Fig1}(e) but dramatically simpler and with fewer parameters than a microscopic model\cite{yang2012strong,romhanyi2014entangled}. The energy of optical modes at the $\Gamma$ point (also the bandwidth of magnon bands below 13 meV) is $8|J_1+J_2|\approx 12$ meV, while the M point splitting reflects the difference between $J_1$ and $J_2$: $4|J_1-J_2|\approx 1.2$ meV. Following the previous DFT calculation \cite{janson2014quantum} and assuming that $|J_1|<|J_2|$ leads to the parameters and calculated magnon dispersion in Fig.~\ref{Fig2}(a) (magenta). High temperature expansion yields \cite{janson2014quantum} $\Theta_{CW}\approx-4(J_1+J_2)=70$ K, which is consistent with the Curie-Weiss temperature  $\Theta_{CW}=69(2)$~K extracted from high temperature susceptibility data \cite{bos2008magnetoelectric}. However, contrary to the helimagnetic state of $\rm Cu_2OSeO_3$, this model is a FM and it does not yet account for the previously enumerated features (Splitting of magnon modes at the $\Gamma$ and $R$ points, broadening of the lower magnon branches at the $X$ point) of the high resolution data in Sec.~\ref{Exp} nor the helical ground state. \par

To account for these, we augment the model with symmetry allowed DM interactions:
\begin{equation}
\mathcal{H}_D = \sum_{\langle ij \rangle}\mathbf{D}_{ij}\cdot(\mathbf{S}_i\times\mathbf{S}_j)
	+\sum_{\langle\langle ij\rangle\rangle}\mathbf{D}'_{ij}\cdot(\mathbf{S}_i\times\mathbf{S}_j).
\end{equation}
The nearest neighbor DM vectors $\mathbf{D}_{ij}$ are related to each other by lattice symmetries and can be expressed in terms of their coordinates in a local frame, $\mathbf{D}_{ij} = (d_1, d_2, d_3)$. The same applies to the second-neighbor DM vectors $\mathbf{D}'_{ij}$.  The absence of mirror symmetries in \cu2 means there are no constraints on these 6 parameters.  The DM vectors for each bond are in Table \ref{conv} of Appendix~\ref{LWST}. The DM vector for a representative nn bond is shown in Fig.~\ref{Fig2}. Determining the exact ground state and spin wave dispersion relation for a general set of DM interactions is non-trivial. Appendix \ref{LWST} describes a semi-quantitative analysis the results of which we shall now summarize. \par
\subsection{R Point Splitting}
 The $R$-point splitting $\Delta_R=1.6(2)$ meV is closely related to DM components $d_1$ and $d_1'$, which mix the magnon modes of the four sublattices in the coarse-grained unit cell. Specifically we find $\Delta_R=4|d_1-d_1'|$. Field theoretical analysis \cite{janson2014quantum} yields the following expression for the helical pitch $|\mathbf k_h|\propto |d_1+d_1'|$ when all other DM components are 0. We note that the splitting at the R point, $\Delta_R$, is independent of whether the ground state is incommensurate (whether $k_h$ is finite). The little group of the lattice space group $P2_13$ at the R point has no four-dimensional irreducible representation to protect any four-fold degeneracy \cite{elcoro2017double} even when the magnetic structure is commensurate. It follows that if $d_1$ and $d_1'$ were the only anisotropy parameters, they would be uniquely determined by $\Delta_R$ and $k_h$. While symmetric anistropic exchange can also contribute to  $\Delta_R$, the absence of a significant $\Gamma$ point gap in the excitation spectrum as indicated by the present data ($\Delta_{\Gamma}\leq 0.1$~meV), microwave \cite{kobets2010microwave} and specific heat \cite{prasai2017ballistic} data however constrain such anisotropy terms.  \par
\subsection{X Point Broadening}
The lower branch of the  $X$ point magnon dispersion should be two-fold degenerate, because the corresponding little group of $P2_13$ only has two-dimensional irreducible representations \cite{elcoro2017double}. For an incommensurate ground state, the symmetry of the magnon Hamiltonian is lowered by the magnetic structure which selects one particular $\langle100\rangle$ direction. Thus the $X$ point along the magnetic wave vector (defined as $Z$) is distinguishable from the orthogonal $X$-directions. Our measurements are however, carried out on a multi-domain sample so that $X$ and $Z$ point data are superimposed. This effect may contribute to the $X$-point broadening though it cannot account for the continuum between 4 and 8 meV at the $X$-point (Fig~\ref{Fig1}(e),Fig~\ref{Fig3}(h-j)).  \par

In Fig.~\ref{Fig2}(a), we also indicate the phase space for two-magnon states. The colormap background indicates areas in $\mathbf{P}-E_2(\mathbf{P})$ space where $\mathbf{P}=\mathbf{p}_1+\mathbf{p}_2$ and $E_2(\mathbf{P})=E(\mathbf{p}_1)+E(\mathbf{p}_2)$ represents the two-magnon continuum for a given momentum $\mathbf P$. Here $E(\mathbf{p}_1)$ is the energy of single magnons given by $\mathcal{H}_J$ with momentum $\mathbf{p}_1$. We notice the shape of the two-magnon continuum near the $X$ point and along the $MR$ edge closely resembles the broadened region of the inelastic neutron data (see Fig.~\ref{Fig1}(e)). This suggests possible 1 to 2 magnon decay allowed by the non-collinear magnetic structure, as observed in various magnetic systems\cite{Stone_2006,Plumb_2015}. The crossing of the single magnon dispersion through the two-magnon phase space means the kinematic constraints (conservation of energy and momentum) are satisfied. This is a necessary but not sufficient condition for spontaneous magnon decay \cite{zhitomirsky2013colloquium}. The lower branch of the magnon modes around the $X$ point can in principle decay into two acoustic magnons. The density of states (DOS) of the two-magnon continuum reflects the number of one- to two-magnon decay channels. However, the resulting line width (decay rate) is controlled by the magnitude of interaction vertices: indeed the single-magnon modes with most significant broadening (the lower modes at the $X$ point and the $XM$ and $XR$ edges) do not coincide with the largest two-magnon continuum DOS. It is interesting to note however, that the observed scattering intensity near the $X$-point closely follows the calculated two magnon continuum. This points to the possibility that single magnon excitations are completely destabilized in this region of the Brillouin zone and replaced by two-magnon excitations. \par

Another possible mechanism for broadening at the said momenta is magnon-phonon interactions. The previous inelastic neutron scattering experiment at $T=70~K$ \cite{tucker2016spin} reported an acoustic phonon mode around 5 meV and an optical phonon around 8 meV at the $X$ point. These two phonons overlap with the broadened lower branches of magnons at the $X$ point and along the $XR$ edge. The hybridization of crossing magnon and phonon modes at the zone boundary may play a role in the apparent magnon decays. A similar explanation was proposed for magnon softening in ferromagnetic manganese perovskites\cite{Dai_2000}. A  thorough quantitative analysis is needed to distinguish between these distinct scenarios.\par
 
 \subsection{Splitting of Optical Modes at the $\Gamma$ Point }
The splitting of the optical modes at the $\Gamma$ point is affected by $d_2,d_2',d_3,d_3'$, but not by $d_1$ or $d_1'$ (Appendix \ref{LWST},\ref{Reliability}).\par
 
 In Fig.~\ref{Fig2}, we show as green lines the  magnon dispersion calculated for $\mathcal{H}_{\text{tot}}\equiv\mathcal{H}_J+\mathcal{H}_D$ with the same $J_1, J_2$ as previously employed,  $d_1=-d_1'=0.2$ meV, and the remaining DM components set to 0. This is a special case ($d_1=-d_1'$), in which the DM interactions cancel and lead to a collinear FM ground state with $k_h=0$. The experimentally observed energy splitting at the $R$ point is $\Delta_R=1.6$ meV. Note the mode splitting along the $XM$, $XR$, and $MR$ edges due to the multi-domain effect. The optical modes at the $\Gamma$ point however, remain  degenerate. By including other components of the DM interaction the dispersion at the $M$ point is modified so the relationship $4|J_1-J_2|\approx 1.2$ meV associated with the experimentally determined $M$-point splitting does not strictly hold in the following numerical fit.

\section{Quantitative comparison}\label{Num}

\begin{figure*}[!htbp] 
	\includegraphics[width=\textwidth]{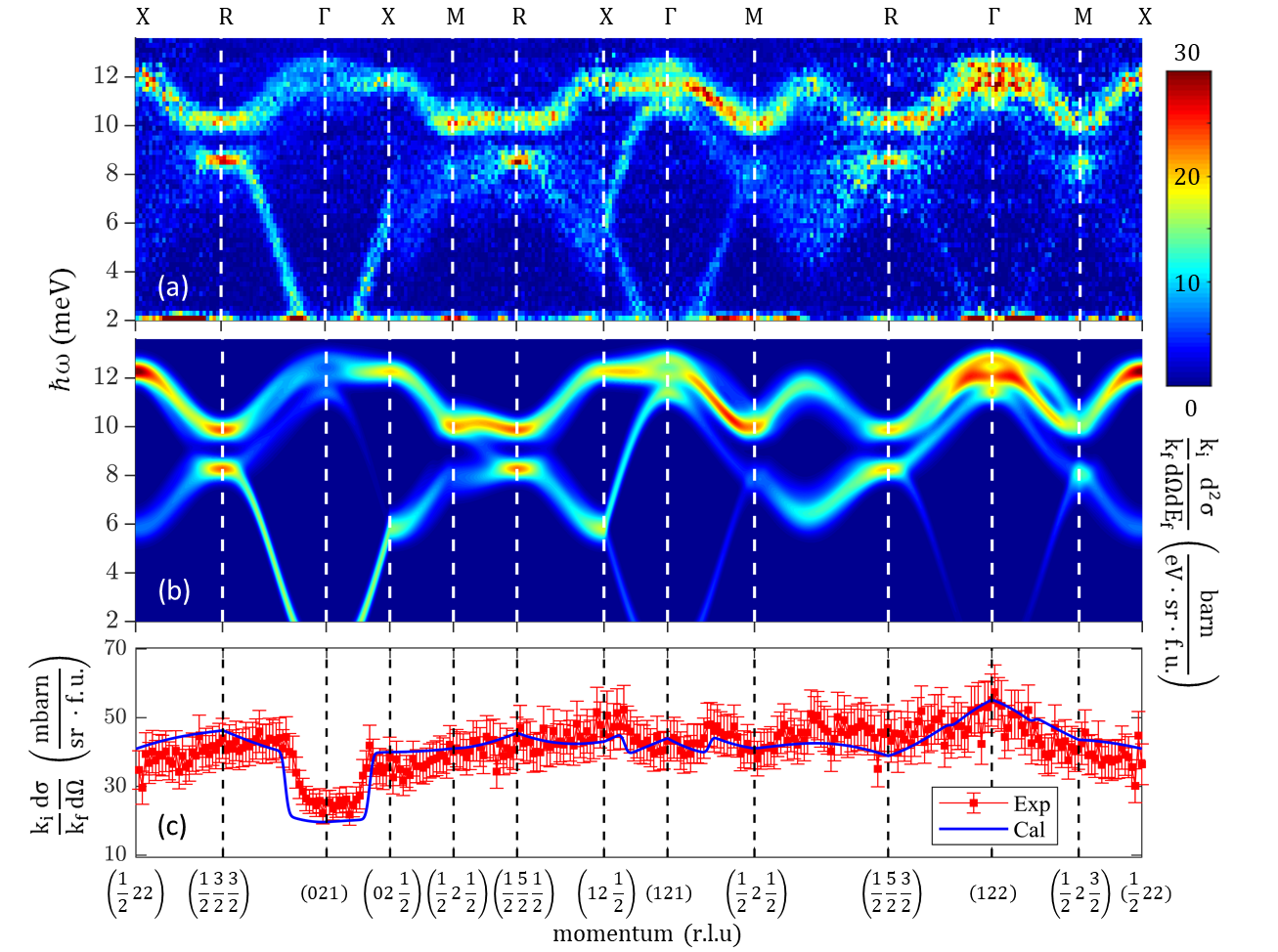}
	\caption{Comparison between experimental (a) and calculated (b) cross section along a path in momentum space that connects labeled high symmetry points. The color bars indicate the intensity scale. In (a), the integration range of perpendicular $\mathbf{Q}$ direction is $\pm 0.05~\mathrm{\AA^{-1}}$. (c) shows the measured and calculated integrated intensity $S(\mathbf{Q})$ (calculated result is multiplied by the constant of proportionality $C$, see Sec.\ref{Dis} (4)). The excellent agreement throughout multiple zones validates the effective-spin formalism and the use of an effective-spin form factor. Error bars in (c) represent one standard deviation.}
	\label{Fig4}
\end{figure*}
To make further progress towards an accurate effective-spin Hamiltonian  $\mathcal{H}_{\text{tot}}$  for $\rm Cu_2OSeO_3$, we use the Matlab Library SpinW\_R3176 \cite{toth2015linear} to calculate the dynamical structure factor for approximate single wavevector helical ground states. Multiple domains are superimposed in our multi-domain sample. Though there exist several theoretical methods to calculate the ground state wavevector and chirality or handedness of the magnetic helicoid from microscopic parameters \cite{janson2014quantum,chizhikov2015microscopic}, in this work we use a numerical approach to obtain the magnetic ground state for a given set of interaction parameters during the optimization of $\mathcal{H}_{\text{tot}}$. First we use the Luttinger-Tisza method\cite{Litvin_1974} to determine the overall magnetic wavevector. We then use the Monte-Carlo method to optimize the relative directions of the 4 effective spins. These steps are repeated until we obtain a single wavevector state with the lowest possible energy. We require the resulting wavevector to be consistent with the magnetic wavevector $k_h$ \cite{adams2012long} and the chirality previously determined by SANS\cite{dyadkin2014chirality}.\par

For comparison with the measured neutron scattering cross section we must take into account the internal  structure of the effective spin. As detailed in Appendix \ref{Formderive}, this is accomplished by multiplying the effective-spin cross section with the formfactor of a ferrimagnetic tetrahedron. The instrumental resolution was handled approximately by replacing the delta-function spectral function of the idealized spin wave cross section with a gaussian energy resolution function. To the calculated $E_i-$dependent energy resolution of the instrument, we added a phenomenological width $2\bar{\Gamma}=0.37$ meV in quadrature to match the experimental FWHM at the $R$ point (see Appendix~\ref{reso} for details). Possible origins of $\bar{\Gamma}$ include a finite spin wave relaxation rate for the gapless non-collinear spin structure and apparent broadening due to the down-folding effects associated with the incommensurate spin structure. The finite $\mathbf{Q}-$resolution of the instrument is not explicitly included and could also in part be the origin of $\bar{\Gamma}$. We then carry out a pixel by pixel least squares fit of the measured versus calculated ${\bf Q}$ and $\hbar\omega$ dependent intensity.  For each set of interaction parameters in $\mathcal{H}_{\text{tot}}$ we determined the constant of proportionality $C$ between model and data by fitting the equal time structure factor $S(\mathbf{Q})=\int_0^{\infty}d\omega S(\mathbf{Q},\omega)$. Two enantiomers  and  three magnetic domains with $\mathbf{k}_h$ along different $\langle100\rangle$ directions were superimposed in the calculated $S(\mathbf{Q},\omega)$. The corresponding measured vs calculated structure factor is shown in Fig.~\ref{Fig4}. For a quantitative examination of the quality of this constrained fit, Fig.~\ref{Fig3} further shows cuts versus energy of $S(\mathbf{Q},\omega)$ at selected high symmetry points in the Brillouin zone. The best-fit parameters thus extracted are listed in Table \ref{Pa}. The calculated dispersion from this set of parameters in the energy range below 1.2 meV is shown in Fig.~\ref{MACS1}(d) to compare with the MACS data shown in Fig.~\ref{MACS1}(c). Resolution effects play a significant role here and are  partially taken into account as described in Sec.~\ref{MRcal}. Momentum space broadening associated with the incommensurate nature of the ground state is also apparent in this low energy regime (Sec.~\ref{MRcal}). Fitting the raw data to an isotropic quadratic dispersion of the form  $E(q)=Dq^2+\Delta_{\Gamma}$ yields $D=67(8)~\text{meV}~\text{\AA}^2$, $\Delta_{\Gamma}=0.0(1)$ meV, slightly larger than the model, which yields  $D=58(2)~ \text{meV}~\text{\AA}^2$ and $\Delta_{\Gamma}=0^{+0.03}_{-0.01}$ meV.  Note that here we are not probing the lower energy regime where helimagnons can be expected for $q<k_h$ and $\hbar
\omega\leq 0.1$ meV. 
  \begin{center}
	\begin {table*}  [!htbp] 
	\begin{tabular}{|c|c|c|c|c|c|c|c|c|} 
		\hline
		Parameter&$J_1$ & $J_2$ & $d_1$ & $d_2$ &$d_3$ &$d_1'$ &$d_2'$ &$d_3'$ \\ [0.5ex] 
		\hline
		Best fit (meV)&$-0.58_{-0.03}^{+0.08}$&$-0.93_{-0.07}^{+0.10}$&	$0.24_{-0.03}^{+0.02}$&	$-0.05$&	$-0.15$&	$-0.16_{-0.03}^{+0.02}$&	$-0.10$&	$0.36$ \\ 
		\hline
	\end{tabular}
	\caption {Optimized parameters resulting from the pixel to pixel fit, shown in Fig.~\ref{Fig3} and Fig.~\ref{Fig4}. These parameters stabilize a helimagnetic ground state with $k_h=0.0143$ r.l.u (compared to $0.0145(11)$ r.l.u from \cite{adams2012long}) and with the same magnetic chirality as the lattice chirality \cite{dyadkin2014chirality}. The range of confidence is given for $J_1,J_2,d_1,d_1'$, there are four sectors of parameters with $J_1,J_2$ and $d_1,d_1'$ interchanged that produce a similar quality fit. $d_2,d_3,d_2',d_3'$ are not well bound in this fit. See Appendix \ref{Reliability} for a more detailed discussion of what can be said about these model parameters based on the neutron data. Specifically, we obtain three empirical constraints on $d_2,d_3,d_2'$, and $d_3'$.}
	\label{Pa}
	\end {table*}
\end{center}

\section{Discussion}\label{Dis}
	Fig.~\ref{Fig3} and Fig.~\ref{Fig4}  show good agreement between model and data both in terms of dispersion and intensity. The effective model $\mathcal{H}_{\text{tot}}$ with only 4 parameters ($J_1,J_2,d_1,d_1'$) already accounts for most of the features of the measured magnon dispersion, including the $R$ point splitting which requires anistropic interactions \cite{tucker2016spin}. Despite playing a secondary role and being less bounded by the measured inelastic neutron scattering data, $d_2,d_2',d_3$ and $d_3'$ are included to account for the the splitting of the optical modes at the $\Gamma$ point and the broadening of peaks at $M$. This shows DM interactions can have a non-negligible influence on magnon spectra beyond the low energy regime, while still stabilizing an incommensurate ground state with small $k_h$ consistent with previously reported SANS data. The consistency of the calculated and measured intensity throughout multiple Brillouin zones validates the use of an effective form factor for cluster-spins and solidifies the hierarchical approach to this compound. Several discrepancies however, remain due to the complexity of the physical system and the limits of the model, which we discuss individually here. \par
	
	(1) Since the ground state is helical and incommensurate, with real space periodicity $\frac{2\pi}{k_h}$, the period of the magnon dispersion should be $k_h$ in the direction of the wavevector instead of 1 r.l.u. For a single magnetic domain with $\mathbf{k}_h$ along certain $\langle100\rangle$ direction, the observable magnon modes at $\mathbf{q}$ with $q_{\perp}\neq 0$ ($q_{\perp}$ is the component of $\mathbf{q}$ perpendicular to $\mathbf{k}_h$) are magnon modes originating from $\Gamma$ points (denoted as $\mathbf{q}$ mode) and those from $\pm N\mathbf{k}_h$ (denoted as $\mathbf{q}\pm N\mathbf{k}_h$ mode with $N\geq 1$). Along the direction of $\mathbf{k}_h$ ($q_{\perp}=0$), we expect to observe only $\mathbf{q}$ and $\mathbf{q}\pm \mathbf{k}_h$ modes if we have a single $k_h$ helical ground state, while the cantings and phase shifts due to multiple sublattices and possible higher-order spin-orbital coupling terms may include additional modes with less weights\cite{Janoschek_2010,Kugler_2015}. In our measured cross-section, due to the presence of multiple magnetic domains, we generally expect to observe $\mathbf{q}\pm N\mathbf{k}_h$  modes at any finite $\mathbf{q}$. For practical reason we only include $\mathbf{q}$ and $\mathbf{q}\pm\mathbf{k}_h$ in the calculation, therefore all high order folding modes are neglected. A $\Gamma$ point magnetic excitation at 8.4 meV was detected by THz optical spectroscopy\cite{laurita2017low}, which also can be observed in our neutron data (see Fig~\ref{Fig3}(a,b)). It was interpreted as a magnon folded back from high momentum. This mode does not appear in our calculation, which is presumably because our model does not properly take into account such down-folding effects. \par
	 
	  (2) The model treats each cluster as a rigid classical spin-1, which is equivalent to assuming $J_s^{\text{AF}}\rightarrow \infty$ when in fact $J_s^{\text{AF}}=12.5$ meV\cite{portnichenko2016magnon} is large but finite. As a result, the ground state will be a superposition of spin-1 and spin-2 states due to exchange interactions with neighboring tetrahedra\cite{romhanyi2014entangled}, as well as of spin-0 states due to intra-tetrahedra DM interactions. The effects of this can be seen in the ratio between the magnon energy at  the $\Gamma$ point and the center of the two modes at the $R$ point. This ratio is strictly 4:3 in the rigid cluster model. In the measured data, the energy of optical modes at the $\Gamma$ point is around 11.6(2) meV so that the model correspondingly would predict a center energy of 8.7(2) meV at the $R$ point.  The center energy at the $R$ point is however observed slightly higher at 9.2(2) meV. This 0.5 meV deviation can not be accommodated in the rigid spin-1 model by varying the exchange parameters. Instead the fit procedure leads to a compromise as in Fig.~\ref{Fig3}. This deviation may also be caused by the magnon-phonon coupling between the two magnon modes and the 6.9 meV phonon mode that we identify in Fig. \ref{Fig3}(e-g) and (k). A similar phonon magnetochiral effect was recently proposed in the context of an ultrasound experiment\cite{Nomura_2019}.
	  
	  (3) The overall broadening of magnon peaks exceeds the instrument resolutions corresponding to a relaxation rate $\bar{\Gamma}=0.18(5)$ meV throughout the Brillouin zone. At the $X$ point between 4 and 8 meV (see Fig.~\ref{Fig3} (h-j)), the single magnon branch actually vanishes and is replaced by continuum scattering in a region of ${\bf Q}-\omega$ space that closely matches that of the kinematically allowed two-magon continuum. The broadenings of the upper magnon branch (around 12 meV) at the $X$ point also exceeds the average phenomenglocal FWHM corresponding to $\bar{\Gamma}$ (see Appendix~\ref{reso}). We believe these effects arise from magnon interactions and decay processes as should be anticipated for a low symmetry and low spin ($S=1$) gapless magnet.\par
	  
	  (4) In this study we have used two methods to normalize the neutron data. The first is vanadium incoherent scattering, which gives a normalization factor $N_v$ with systematic uncertainty $\approx 15\%$. We further calculate and compare the Bragg intensities (Appendix~\ref{Bn}), and get a normalization factor $N_B\approx 1.2N_v$ with $\approx 30\%$ uncertainty. Throughout the paper we have adopted $N_B$ for data normalization. The constant of proportionality $C$ (ratio) between normalized measured magnetic cross section and calculated cross section is fitted to be $1.15(5)$. Considering the presence of phonon cross-sections and background scattering, the calculated result of our rigid spin-cluster model is consistent with the experimental data normalized by $N_B$ within uncertainty. Besides limitations in the accuracy of the absolute normalization of the measured neutron scattering cross section, the following reasons may also cause discrepancy between calculated and measured magnetic cross-section: (1) The spin density distribution around $\text{Cu}^{2+}$ may be more extended than for atomic $3d^9$ electrons\cite{dianoux2002neutron}, even spreading onto the ligand sites. This may cause a more rapid decrease of the magnetic form factor $F(\mathbf{Q})$ (see Appendix.~\ref{Formderive}) as a function of $Q$ than accounted for in the analysis. (2) The ground state and low energy excited states of the system may be more entangled\cite{ozerov2014establishing,romhanyi2014entangled} than the rigid limit we take. Such quantum entanglement may reduce (increase) the effective spin length for each $\text{Cu}^{2+}$ by admixing spin-0 (spin-2) states into the ground state and the low energy excited states. (3) The high order folding modes ($\mathbf{q}\pm N\mathbf{k}_h$, $N>1$) we neglect may cause the distribution of spectral weights to differ from calculations neglecting these components. (4) Furthermore, the finite momentum resolution of the instrument has not been fully quantified and included in the comparison between model and data.
	
	\section{Conclusion}\label{Concl}
	\cu2 is a complex low symmetry magnetic material. The complexity starts with a large structural unit cell containing 16 magnetic ions. The lack of inversion symmetry gives rise to a chiral magnetic order with a periodicity that is incommensurate with the crystalline lattice. Understanding the spectrum of excitation in such a magnet is a non-trivial task that we dedicated ourselves to in this paper. 
	
	We conducted an inelastic neutron scattering experiment on \cu2 focusing on the 4 lowest magnon branches and built a quantitative effective spin model that can be the basis for describing its low energy magnetism. The model includes DM interactions that stabilize the helimagnetic order. Features of the magnon spectrum missed in previous experiments and calculations have been quantitively established and related to the incommensurability of the magnetic order. The interaction parameters were obtained by fitting the model to $\mathbf{Q}-E$ slices through four dimensional inelastic magnetic neutron scattering data. The resulting coarse-grained model provides an accurate description of the four lowest energy branches of the magnon spectrum. The methods exemplified by this work can be extended to other magnets where dominant interactions lead to the formation of effective spins at low energies. Our model will facilitate  understanding of the complicated phase diagram of \cu2 including the exotic skyrmion phase.
	
	\section*{Acknowledgments}
	This work was supported as part of the Institute for Quantum Matter, an Energy Frontier Research Center funded by the U.S. Department of Energy, Office of Science, Basic Energy Sciences under Award No. DE-SC0019331. CB and JK were supported by the Gordon and Betty Moore foundation under the EPIQS program grant number GBMF-4532. Access to MACS was provided by the Center for High Resolution Neutron Scattering, a partnership between the National Institute of Standards and Technology and the National Science Foundation under Agreement No. DMR-1508249.We wish to thank Jonathan Gaudet and Predrag Nikolic for the useful discussion on understanding the $X$-point broadening, and Jiao Lin for helping evaluating the instrumental resolution of SEQUOIA.

\appendix
\setcounter{table}{0}\renewcommand\thetable{\Alph{table}}
\setcounter{figure}{0}\renewcommand\thefigure{\Alph{section}\arabic{figure}}
\renewcommand{\thesubsubsection}{\alph{subsubsection}}

\section{Details of the spinwave model}\label{LWST}
\begin{table*}[!htbp] 
	\captionsetup[subfloat]{justification=centering}
	\centering
	\subfloat[Subtable 1 list of tables text][Unit cell of 16 $\mathrm{Cu^{2+}}$.]{
	\includegraphics[width=0.5\columnwidth]{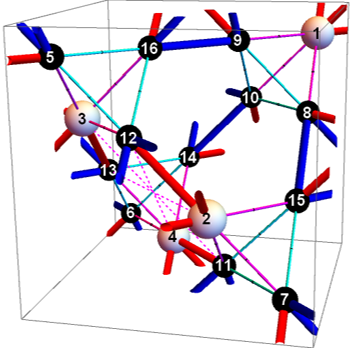}}
	\qquad
	\subfloat[Subtable 2 list of tables text][Coordinate Table, the same as in Ref.~\cite{janson2014quantum}]{
	\begin{tabular}[b]{|c|c|c|c|}\hline
	Labels & Coordinates &Labels & Coordinates\\ \hline
	$\mathbf{\rho}_1$&$(y,y,y)$ &$\mathbf{\rho}_2$&$(\frac{3}{2}-y,1-y,y-\frac{1}{2})$	\\
	\hline
	$\mathbf{\rho}_3$&$(1-y,y-\frac{1}{2},\frac{3}{2}-y)$ &$\mathbf{\rho}_4$&$(y-\frac{1}{2},\frac{3}{2}-y,1-y)$\\
		\hline
	$\mathbf{\rho}_5$&$(a,b,c)$ &$\mathbf{\rho}_6$&$(b,c,a)$\\	\hline
	$\mathbf{\rho}_7$&$(c,a,b)$&$\mathbf{\rho}_8$&$(1-a,b+\frac{1}{2},\frac{3}{2}-c)$\\	\hline
	$\mathbf{\rho}_9$&$(b+\frac{1}{2},\frac{3}{2}-c,1-a)$ &$\mathbf{\rho}_{10}$&$(\frac{3}{2}-c,1-a,b+\frac{1}{2})$\\	\hline
	$\mathbf{\rho}_{11}$&$(a+\frac{1}{2},\frac{1}{2}-b,1-c)$ &$\mathbf{\rho}_{12}$&$(\frac{1}{2}-b,1-c,a+\frac{1}{2})$\\	\hline
	$\mathbf{\rho}_{13}$&$(1-c,a+\frac{1}{2},\frac{1}{2}-b)$ &$\mathbf{\rho}_{14}$&$(\frac{1}{2}-a,1-b,c-\frac{1}{2})$\\	\hline
	$\mathbf{\rho}_{15}$&$(1-b,c-\frac{1}{2},\frac{1}{2}-a)$ &$\mathbf{\rho}_{16}$&$(c-\frac{1}{2},\frac{1}{2}-a,1-b)$\\	\hline
		\end{tabular}}
	\caption{Coordinates of 16 $\text{Cu}^{2+}$ sites in the unit cell of a right-handed enantiomer, where $y=0.88557$, $a=0.13479$ (not to be confused with lattice constant), $b=0.12096$, and $c=0.87267$. The unit cell is plotted in (a), where Cu-1 (white), Cu-2 (black), $J_s^{\text{AF}}$ (red), $J_s^{\text{FM}}$ (blue), $J_w^{\text{AF}}$ (magenta),$J_w^{\text{FM}}$ (cyan) and $J_{\text{o.o}}^{\text{AF}}$ (magenta,dashed) are plotted with the same convention as in Fig.~\ref{Fig1}(a).We use the position of  Cu-1 ($\mathbf{\rho}_1$ to $\mathbf{\rho}_4$ to represent the position of each cluster.}
\end{table*}\label{Posit}

In the main text, we consider interactions between nn and nnn clusters. All $J_{ij}$ and $\mathbf{D}_{ij}$ are listed in Table \ref{conv}. Through gradient expansion and field theory analysis previous studies indicated a single wavevector helical state\cite{janson2014quantum} at zero field and low temperature. However, due to the rather low lattice symmetry, the exact ground state will also involve canting and phase shifts among the 4 sublattices. This raises significant difficulty in analytically determining the exact magnetic structure with a general set of DM interactions. Furthermore, the helical modulation mixes spinwave modes with momentum $\mathbf{q}$ and $\mathbf{q}\pm N\mathbf{k}_h$, with $N=1,2,...$ and $\mathbf{k}_h$ is the helical state wave vector. In the following we will only consider mixings between $\mathbf{q}$ and $\mathbf{q}\pm\mathbf{k}_h$ modes.\par

\begin{table*}[!htbp] 
	\begin{center}
		\begin{tabular}{c c} 
			\begin{tabular}{|c|c|c|c|} 
				\hline
				$\mathbf{r}_i$  &$\mathbf{r}_j$ & $J_{ij}$& $\mathbf{D}_{ij}$ \\ 
				\hline
				
				$\mathbf{\rho}_1$&$\mathbf{\rho}_4+[0,0,1]$&$J_1$&$(d_1,d_2,d_3)$\\
				\hline
				$\mathbf{\rho}_1$&$\mathbf{\rho}_2+[0,1,0]$&$J_1$&$(d_2,d_3,d_1)$\\
				\hline
				$\mathbf{\rho}_1$&$\mathbf{\rho}_2+[0,1,1]$&$J_1$&$(d_2,d_3,-d_1)$\\
				\hline
				$\mathbf{\rho}_1$&$\mathbf{\rho}_3+[1,0,0]$&$J_1$&$(d_3,d_1,d_2)$\\
				\hline
				$\mathbf{\rho}_1$&$\mathbf{\rho}_4+[1,0,1]$&$J_1$&$(-d_1,d_2,d_3)$\\
				\hline
				$\mathbf{\rho}_1$&$\mathbf{\rho}_3+[1,1,0]$&$J_1$&$(d_3,-d_1,d_2)$\\
				\hline
				$\mathbf{\rho}_2$&$\mathbf{\rho}_4+[0,-1,0]$&$J_1$&$(-d_3,d_1,d_2)$\\
				\hline
				$\mathbf{\rho}_2$&$\mathbf{\rho}_3$&$J_1$&$(d_1,-d_2,d_3)$\\
				\hline
				$\mathbf{\rho}_2$&$\mathbf{\rho}_4$&$J_1$&$(-d_3,-d_1,d_2)$\\
				\hline
				$\mathbf{\rho}_2$&$\mathbf{\rho}_3+[1,0,0]$&$J_1$&$(-d_1,-d_2,d_3)$\\
				\hline
				$\mathbf{\rho}_3$&$\mathbf{\rho}_4$&$J_1$&$(-d_2,d_3,d_1)$\\
				\hline
				$\mathbf{\rho}_3$&$\mathbf{\rho}_4+[0,0,1]$&$J_1$&$(-d_2,d_3,-d_1)$\\
				\hline
			\end{tabular}
			&
			\begin{tabular}{|c|c|c|c|} 
				\hline
				$\mathbf{r}_i$  &$\mathbf{r}_j$ & $J_{ij}$& $\mathbf{D'}_{ij}$ \\ 
				\hline
				$\mathbf{\rho}_1$&$\mathbf{\rho}_4+[0,1,1]$&$J_2$&$(d_1',d_2',d_3')$	\\
				\hline
				$\mathbf{\rho}_1$&$\mathbf{\rho}_3+[1,0,1]$&$J_2$&$(d_3',d_1',d_2')$	\\
				\hline
				$\mathbf{\rho}_1$&$\mathbf{\rho}_2+[1,1,0]$&$J_2$&$(d_2',d_3',d_1')$	\\
				\hline
				$\mathbf{\rho}_1$&$\mathbf{\rho}_2+[1,1,1]$&$J_2$&$(d_2',d_3',-d_1')$	\\
				\hline
				$\mathbf{\rho}_1$&$\mathbf{\rho}_3+[1,1,1]$&$J_2$&$(d_3',-d_1',d_2')$	\\
				\hline
				$\mathbf{\rho}_1$&$\mathbf{\rho}_4+[1,1,1]$&$J_2$&$(-d_1',d_2',d_3')$	\\
				\hline
				$\mathbf{\rho}_2$&$\mathbf{\rho}_3+[0,-1,0]$&$J_2$&$(d_1',-d_2',d_3')$	\\
				\hline
				$\mathbf{\rho}_2$&$\mathbf{\rho}_4+[0,-1,1]$&$J_2$&$(-d_3',d_1',d_2')$	\\
				\hline
				$\mathbf{\rho}_2$&$\mathbf{\rho}_4+[0,0,1]$&$J_2$&$(-d_3',-d_1',d_2')$	\\
				\hline
				$\mathbf{\rho}_2$&$\mathbf{\rho}_3+[1,-1,0]$&$J_2$&$(-d_1',-d_2',d_3')$	\\
				\hline
				$\mathbf{\rho}_3$&$\mathbf{\rho}_4+[-1,0,0]$&$J_2$&$(-d_2',d_3',d_1')$	\\
				\hline
				$\mathbf{\rho}_3$&$\mathbf{\rho}_4+[-1,0,1]$&$J_2$&$(-d_2',d_3',-d_1')$	\\
				\hline
			\end{tabular}
		\end{tabular}
	\end{center}
	\caption{Conventions for the nn and nnn bonds. $\mathbf{r}_i$ and  $\mathbf{r}_j+[m,n,p]$ label the coordinates of clusters (strong tetrahedra) in units of the cubic lattice parameter. $J_{ij}$ and  $\mathbf{D}_{ij}$ are the Heisenberg and DM interactions between site $i$ and $j$. We choose the same convention as listed in Table 3 of \cite{janson2014quantum}.}
\end{table*}\label{conv}

\subsection{$R$ point splitting}\label{Rgap}
The $R$ point splitting can be related to two specific DM components listed in Table \ref{conv}, namely $d_1$ for nn and $d_1'$ for nnn. The reason we have a 4-fold degeneracy for the Heisenberg model $\mathcal{H}_j$ is partly due to the symmetry of our coarse-grained lattice structure: we have four sublattices in the unit cell, $\rho_1$ to $\rho_4$ in Table \ref{Posit}. Each sublattice has 6 nn and 6 nnn. For example, sublattice-1($\rho_1$)  has 2 nn and 2 nnn on each of the sublattice-2, 3 and 4 respectively. Defining $\hat{\mathbf{e}}_3$ to be the global direction of magnetization for the collinear ground state, while $\hat{\mathbf{e}}_1$ and $\hat{\mathbf{e}}_2$ are the two orthogonal directions ($\hat{\mathbf{e}}_1$, $\hat{\mathbf{e}}_2$, $\hat{\mathbf{e}}_3$ are chosen to form a right-handed local frame), we consider small deviations from the ground state magnetic structure
\begin{equation}\label{linear}
\delta\mathbf{S}_i=\alpha_i\hat{\mathbf{e}}_1+\beta_i\hat{\mathbf{e}}_2+\left(1-\frac{\alpha^2_i+\beta^2_i}{2}\right)\hat{\mathbf{e}}_3
\end{equation}
where $i$ labels the sublattice, and $\delta\mathbf{S}$, $\alpha$ and $\beta$ are functions of $(n_x,n_y,n_z)$ (labeling the unit cell). Then the magnon dispersion comes from the quadratic terms in $\alpha_i,\beta_i$ within a Taylor expansion of the exchange energy. For a certain sublattice-1, the change in exchange energy resulting from a deviation in spin from the ground state configuration can be written as
\begin{align}
\langle \delta {\cal H}_J\rangle_{1i} &=J_1\delta\mathbf{S}_1\cdot\left(\sum_{i\in nn}\delta\mathbf{S}_i\right)\nonumber\\&+J_2\delta\mathbf{S}_1\cdot\left(\sum_{i\in nnn}\delta\mathbf{S}_i\right)
\end{align}
The definition of the $R$ point $(\frac{1}{2},\frac{1}{2},\frac{1}{2})$ in momentum space is that in real space we have
\begin{align}
\alpha_i(n_x,n_y,n_z)&=(-1)^{n_x+n_y+n_z}\alpha_{i0}\label{R}\\ \beta_i(n_x,n_y,n_z)&=(-1)^{n_x+n_y+n_z}\beta_{i0} \nonumber
\end{align} 
in other words $\delta \mathbf{S}_i$ (to linear order) change signs from one unit cell $(n_x,n_y,n_z)$ to its neighbor ($(n_x\pm 1,n_y,n_z)$, etc). 
Consider the nn terms between sublattice-1 and sublattice-4, 
\begin{align}\label{J14}
&J_1\delta\mathbf{S}_1(n_x,n_y,n_z)\cdot\left[\delta\mathbf{S}_4(n_x,n_y,n_z+1)+\right.\\&\left.\delta\mathbf{S}_4(n_x+1,n_y,n_z+1)\right]\nonumber
\end{align}
The first term (to quadratic order in $\alpha,\beta$) reads
\begin{align}
&J_1\left[\alpha_1(n_x,n_y,n_z)\hat{\mathbf{e}}_1+\beta_1(n_x,n_y,n_z)\hat{\mathbf{e}}_2\right.\\&\left.+\left(1-\frac{\alpha^2_1+\beta^2_1}{2}\right)\hat{\mathbf{e}}_3\right]\cdot\left(\alpha_4(n_x,n_y,n_z+1)\hat{\mathbf{e}}_1\right.\nonumber\\&\left.+\beta_4(n_x,n_y,n_z+1)\hat{\mathbf{e}}_2+\left(1-\frac{\alpha^2_4+\beta^2_4}{2}\right)\hat{\mathbf{e}}_3\right)\nonumber
\\\approx&J_1\left(-\alpha_{10}\alpha_{40}-\beta_{10}\beta_{40}-\frac{\alpha^2_{10}+\beta^2_{10}}{2}-\frac{\alpha^2_{40}+\beta^2_{40}}{2}\right)\nonumber
\end{align}
Only the first two terms involve interactions between different modes and can split the degeneracy, however, sublattice-1 has another nn of sublattice 4 (the second term in \ref{J14}) which is exactly one unit cell away, which contributes quadratic terms as
\begin{align}
J_1\left(\alpha_{10}\alpha_{40}+\beta_{10}\beta_{40}-\frac{\alpha^2_{10}+\beta^2_{10}}{2}-\frac{\alpha^2_{40}+\beta^2_{40}}{2}\right)
\end{align}
and exactly cancels the cross-terms between sublattice-1 and sublattice 4. A similar cancellation occurs between all other sublattices and again for nnn terms. The absence of cross-terms between 4 modes leads to a 4-fold degeneracy, even though the cubic group has no 4-dimensional irreducible representation. The splitting at the $R$ point then becomes susceptible to the normally weaker anistropic interactions.\par

Strictly speaking, after turning on DM interactions, we will have a non-collinear ground state. Furthermore the symmetry of the magnon hopping model, determined by the underlying magnetic structure, will be lowered by the helical wavevector selecting a specific [100] direction. We can still estimate the impacts of DM interactions following the above logic. The leading effect of DM interaction (between a certain sublattice-1 and its nn sublattice-4) in the magnon Hamiltoian can be written as follows
\begin{align}
\langle \delta {\cal H}_D\rangle_{1i}&=\mathbf{D}_{14}(0,0,1)\cdot[\delta\mathbf{S}_1(n_x,n_y,n_z)	\times \nonumber\\&\delta\mathbf{S}_4(n_x,n_y,n_z+1)]\nonumber \\&+	 \mathbf{D}_{14}(1,0,1)\cdot[\delta\mathbf{S}_1(n_x,n_y,n_z)\times\nonumber\\&\delta\mathbf{S}_4(n_x+1,n_y,n_z+1)]\label{D14}
\end{align}
where $\mathbf{D}_{14}(0,0,1)=(d_1,d_2,d_3)$ and $\mathbf{D}_{14}(1,0,1)=(-d_1,d_2,d_3)$ can be read from Table \ref{conv}. For a crude estimate, we assume that the ground state is still fairly collinear so we can  still use Eqn.~\ref{linear} and \ref{R}  at the $R$ point. This corresponds to ignoring both the spatial variation of $\hat{\mathbf{e}}_{i}$ and the fact that magnon mode at the $R$ point will naturally mix with those at $(\frac{1}{2},\frac{1}{2},\frac{1}{2})\pm N\mathbf{k}_h$. In other words, since the wavevector $\mathbf{k}_h$ measured in the experiment is quite small, we assume the magnon disperion corresponding to the actual incommensurate ground state can be "adibatically" evolved from some commensurate ground state. In this approximation, we have $\delta\mathbf{S}_4(n_x,n_y,n_z+1)\approx-\delta\mathbf{S}_4(n_x+1,n_y,n_z+1)$ so expression \ref{D14} then reads
\begin{align}
&[\mathbf{D}_{14}(0,0,1)-\mathbf{D}_{14}(1,0,1)]\cdot(\delta\mathbf{S}_1\times\delta\mathbf{S}_4)\\=&2(d_1,0,0)\cdot(\delta\mathbf{S}_1\times\delta\mathbf{S}_4)\propto d_1(\alpha_{10}\beta_{40}-\alpha_{40}\beta_{10})\nonumber
\end{align}
We conclude that since $d_1$ is the only DM component that survives the summation over nn sublattices of the same type, it will predominantly contibute to lifting the degeneracy at the $R$ point by mixing the magnon modes of the four sublattices in the coarse-grained unit cell. The same argument goes for the nnn DM component $d_1'$. A similar argument works for the $\Gamma$ point, where we have $\delta\mathbf{S}_4(n_x,n_y,n_z+1)\approx \delta\mathbf{S}_4(n_x+1,n_y,n_z+1)$, the addition of DM terms contains only $d_2,d_3$ for nn ($d_2',d_3'$ for nnn). Later we will see from numerical calculation that $d_2,d_3,d_2',d_3'$ play major roles in lifting the degeneracy of optical modes at the $\Gamma$ point.\par

We proceed to provide a more quantitive calculation,  that holds when only$J_1$, $d_1$ and $d_1'$ are  non-zero, this is one of the few cases where we can determine the ground state analytically. We use the classical picture, assuming the ground state wavevector is $\mathbf{k}=(0,0,k)$, the ground state configuration is
\begin{align}
\hat{\mathbf{e}}_3(m,\mathbf{n})&=(\cos(\mathbf{k}\cdot\mathbf{r}_{\mathbf{n},m}),\sin(\mathbf{k}\cdot\mathbf{r}_{\mathbf{n},m}),0)\label{J1_GS}\\
\mathbf{r}_{\mathbf{n},m}&=(\mathbf{n}+\left(0,0,\nu_m\right))a\nonumber\\
\nu_1&=\frac{7}{8}\quad\nu_2=\frac{3}{8}\quad\nu_3=\frac{5}{8}\quad\nu_4=\frac{1}{8}\nonumber
\end{align}
here $\hat{\mathbf{e}}_3(m,\mathbf{n})$ represents the direction of magnetization of sublattice-$m$ ($m=1,2,3,4$) in the unit cell labeled by $\mathbf{n}=(n_x,n_y,n_z)$. Substitute \ref{J1_GS} (and similar expressions for $\hat{\mathbf{e}}_1$ and $\hat{\mathbf{e}}_2$) into \ref{linear} and then into the Hamiltonian we obtain the zeroth order expression for the ground state energy 
\begin{align}
f_0=8J_1\cos\left(\frac{ka}{4}\right)+4J_1\cos\left(\frac{ka}{2}\right)-4(d_1+d_1')\sin\left(\frac{ka}{2}\right).
\end{align}
The first order in $\alpha_i$,$\beta_i$ correction vanishes which signals the correct ground state. The wavevector $k$ can be determined by minimizing $f_0$ with respect to $k$, which gives $k\approx -\frac{4(d_1+d_1')}{3J_1a}$. The quadratic in $\alpha_i$,$\beta_i$ energy correction $\mathcal{H}_2$ is too cumbersome to show in full form. For the $R$ point, we consider the mixture between $(\frac{1}{2},\frac{1}{2},\frac{1}{2})$ and $(\frac{1}{2},\frac{1}{2},\frac{1}{2})\pm\mathbf{k}$, which amounts to expanding
\begin{align}
\alpha_i(n_x,n_y,n_z)&=(-1)^{n_x+n_y+n_z}(\alpha_{i0}+\alpha_{i1}\cos(kn_za)\nonumber\\&+\alpha_{i2}\sin(kn_za))\nonumber\\
\beta_i(n_x,n_y,n_z)&=(-1)^{n_x+n_y+n_z}(\beta_{i0}+\beta_{i1}\cos(kn_za)\nonumber\\&+\beta_{i2}\sin(kn_za))\label{Be}
\end{align}
We then substitute the above equations into $\mathcal{H}_2$, integrate out the terms slowly varying in space (terms depending on $n_z$) and only keep leading order terms in $k$. For the Berry phase terms $\dot{\alpha}_i\beta_i$ (see Eqn.~\ref{Be}), terms like $\cos^2(kn_za)$ or $\sin^2(kn_za)$ will give $\frac{1}{2}$ after averaging over spatial regions in z-direction, while crossing-terms with $\sin(kn_za)\cos(kn_za)$ will vanish. By solving the equations of motion for the Lagrangian
\begin{align}
\mathcal{L}=\sum_{i=1}^{4}\left[\dot{\alpha}_{i0}\beta_{i0}+\frac{1}{2}(\dot{\alpha}_{i1}\beta_{i1}+\dot{\alpha}_{i2}\beta_{i2})\right]-\mathcal{H}_2
\end{align}
we find the magnon dispersion energy at the $R$ point to be $\hbar\omega_R=-6J_1\pm2(d_1-d_1')$. That is, the splitting at the $R$ point, $\Delta_R$ ,is approximately $4|d_1-d_1'|$. We can see that the splitting at the $R$ point and the wavevector $\mathbf{k}$, although both related to the microscopic DM interactions, are algebraically independent, $d_1$ and $d_1'$ can be similar in strength to the Heisenberg exchange, while maintaining a small ground state wavevector $\mathbf{k}$ as measured in the experiment.\par

Unfortunately, after including $J_2$, an exact analytical expression for the ground state spin configuration like \ref{J1_GS} is no longer possible. However, in the special case where $d_1=-d_1'$, the effect of nn and nnn DM interactionw exactly cancel in the expression for $k\propto |d_1+d_1'|=0$, resulting in a ferromagnetic ground state where the uniform magnetization can point along any direction. The spinwave dispersion for this case is shown in Fig.~\ref{Fig2}. The corresponding splitting at the $R$ point equals $4|d_1-d_1'|=8|d_1|$.\par
\section{Normalization of neutron data}\label{Bn}
\begin{figure} [!htbp] 
	\includegraphics[width=\columnwidth]{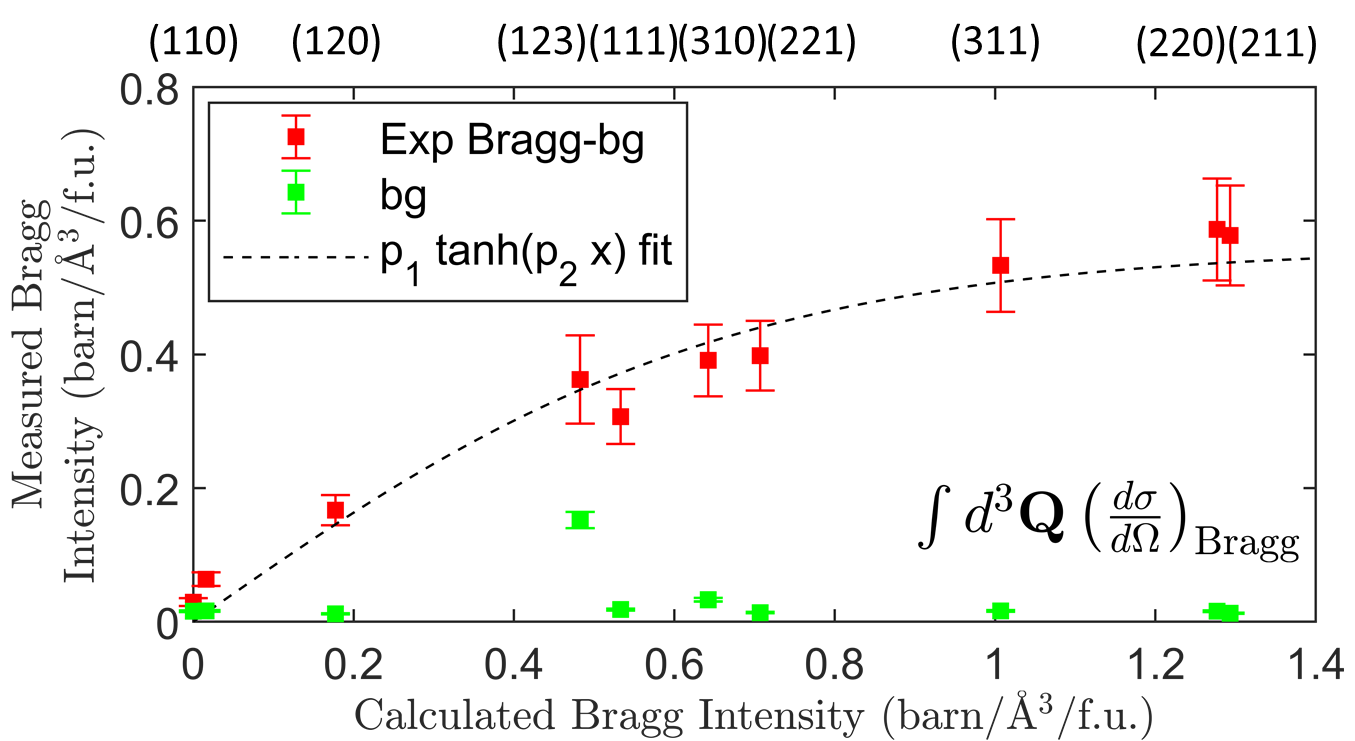}
	\caption{The vanadium normalized experimental $\bf Q$-integrated Bragg  intensities versus calculated nuclear+magnetic Bragg intensities. For each peak a background (green symbols) was subtracted. It was obtained from data acquired at the same $|\mathbf{Q}|$ but with the sample rotated so as to not satisfy the Bragg diffraction. The large background at (123) is due to  powder Bragg diffraction from the aluminum sample mount. The solid line is an empirical fit describing the cross over from a linear to a saturated regime as described in the text. Error bars in the figure represent one standard deviation.}
	\label{bn}
\end{figure}

To check the vanadium normalization, we analyze the $\bf Q$-integrated  intensity of a set of  Bragg peaks. Fig.~\ref{bn} shows the experimental $\bf Q$-integrated Bragg  intensities versus the calculated nuclear+magnetic Bragg intensities. We use an empirical functional form $y=p_1\tanh(p_2x)$ to describethe cross-over from a linear regime for weak Bragg peaks to a saturation regime for strong peaks due to extinction and detector saturation  effects\cite{Hamilton_1958}. The revised normalization factor $N_B=N_v/(p_1p_2)=1.2N_v$, where $N_v$ is the normalization factor inferred from vanadium normalization, indicates 20\% less scattering from the sample than anticipated from the count rates obtained for the vanadium standard sample. While this discrepancy is within systematic errors, we  adopt the Bragg normalization factor $N_B$ as it gauges the  same sample volume and beam area as the inelastic magnetic neutron scattering experiment. 

\section{Derivation of Effective Form Factor}\label{Formderive}
Here we derive the effective form factor of the effective ferrimagnetic spin cluster. The inelastic neutron scattering cross-section measures the transverse spin-spin correlation function, which reads
\begin{align}
&\frac{d^2\sigma}{d\Omega dE_f}(\mathbf{Q},\omega)=\frac{k_f}{k_i}(\gamma r_0)^2\sum_{\alpha,\beta}\left(\delta_{\alpha\beta}-\hat{Q}_{\alpha}\hat{Q}_{\beta}\right)\\&\times\sum_{\mathrm{l,d,l',d'}}\sum_{\mathrm{n,n'}}\left(F_{\mathrm{d'n'}}(\mathbf{Q}) e^{i\mathbf{Q}\cdot\mathbf{r}_\mathrm{{l'd'n'}}}\right)^*\left(F_{\mathrm{dn}}(\mathbf{Q})e^{i\mathbf{Q}\cdot\mathbf{r}_{\mathrm{ldn}}}\right)\nonumber\\&\times\sum_{\lambda,\lambda'}p_{\lambda}\bra{\lambda}
s^{\alpha}_{\mathrm{l'd'n'}}\ket{\lambda'}\bra{\lambda'}s_{\mathrm{ldn}}^{\beta}\ket{\lambda}\delta(E_{\lambda}-E_{\lambda'}+\hbar\omega)\nonumber
\end{align}
$\hbar\omega,\mathbf{Q}$ are energy and momentum transfers, respectively. $k_f,k_i$ are the momentum of final and incoming neutrons, respectively. $\gamma=1.913$, $r_0=2.818\times 10^{-15}$ m is the classical electron radius. We label the spin-$\frac{1}{2}$ of $\text{Cu}^{2+}$ with three indices: $l$ for unit cell, $d=1,2,3,4$ labels the tetrahedral clusters, $n=1,2,3,4$ each $\text{Cu}^{2+}$ within a cluster with $n=1$ corresponding to Cu-1. $F_{dn}(\mathbf{Q})$ is the magnetic form factor of the $\text{Cu}^{2+}$ ion. $p_{\lambda}$ is the probablity that the inital state is $\ket{\lambda}$ with energy $E_\lambda$. The final state $\ket{\lambda'}$ has energy $E_{\lambda'}$. \par
\begin{figure*} [!htbp]
	\includegraphics[width=\textwidth]{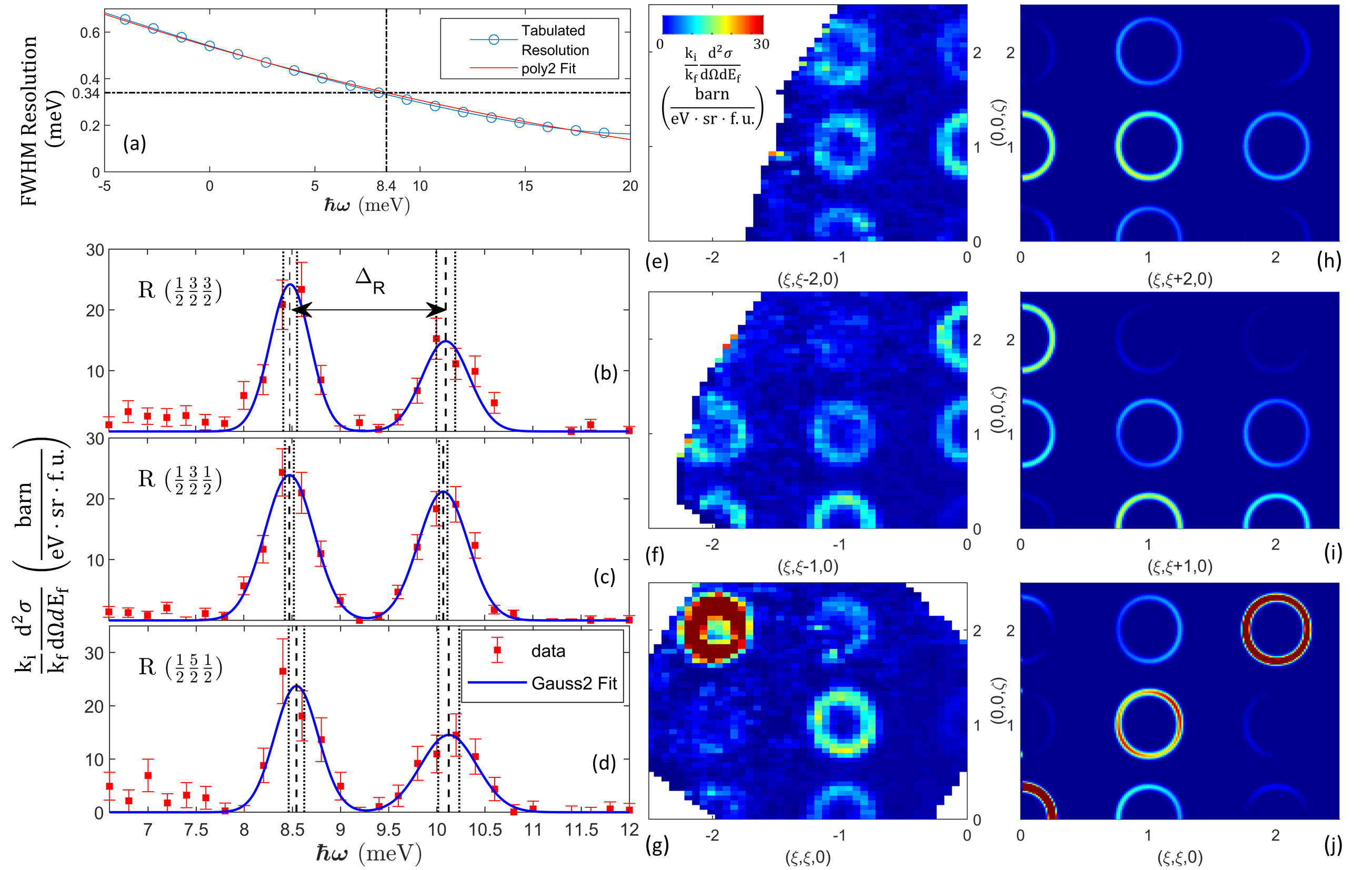}
	\caption{(a) $\hbar\omega$-dependent instrument resolution function of SEQUOIA determined by Monte-Carlo simulation. (b-d) The 2-gaussian fit to energy cuts at different $R$ points. A weak third modes near 6.9 meV (discussed in Sec.~\ref{Exp}) is not considered. Vertical dashed lines indicate the center of the gaussian peak, the dotted lines deliniate  95\% peak position confidence intervals. The splitting at the $R$ point $\Delta_R$ is fitted to be $1.6(2)$  meV. Red symbols show  neutron intensity data averaged over $(0.1~\text{\AA}^{-1})^3\times(0.2~\text{meV})$ in the 4D $\mathbf{Q}-\hbar\omega$ space. (e-g) Constant energy slices averaging over $\hbar\omega \in [2.75, 3.25]$  meV and a range of $\pm 0.1~\text{\AA}^{-1}$ along the $(1\bar{1}0)$ direction. (h-j) Corresponding constant energy slices calculated using SpinW\_R3176, integrating over $[2.7, 3.3]$ meV for equivalent momenta. The agreement between neutron data and model validates the effective form factor Epn.~\ref{Form}. Error bars in (b-d) represent one standard deviation.}
	\label{app2}
\end{figure*}
To proceed we make two key approximations: (1) We take the tabulated form factor\cite{dianoux2002neutron} of $\text{Cu}^{2+}$ for both Cu-1 and Cu-2 (the same for all $F_{\mathrm{dn}}$), that is, we neglect possible influence of the neighbor ligands on spin density distribution of $\text{Cu}^{2+}$ ions. (2) For the magnetic excitation with $\hbar\omega<13$ meV, we take the approximation that $J_s^{\text{FM}},J_s^{\text{AF}}\rightarrow \infty$. In this case, $\ket{\lambda'}$ only includes states wherein each cluster forms an effective spin-1 state, that is, all states $\ket{\lambda'}$ and $\ket{\lambda}$ can be written as direct product states $\ket{S=1,\mathbf{\Omega}}$ for each cluster. Here $\mathbf{\Omega}$ is the angle representing the spin orientation. Any $\ket{S=1,\mathbf{\Omega}}$ can be written as a linear combination of  $\ket{S=1,S_z=0,\pm 1}$, which in turn can be written as a linear combination of $\text{Cu}^{2+}$ states (i.e. $\ket{\uparrow\downarrow\downarrow\uparrow}$, we take the result from Ref.~\cite{romhanyi2014entangled}). For each spin-1, the cluster spin operator reads $S_{\mathrm{ld}}^{\alpha}\equiv\sum_{m=1}^4s^{\alpha}_{\mathrm{ldm}}$. It is then straightforward to work out the relationship between matrix elements $\bra{1,a}s^{\alpha}_{\mathrm{ldm}}\ket{1,b}$ and $\bra{1,a}S^{\alpha}_{\mathrm{ld}}\ket{1,b}$, which can be expressed as follows
\begin{align}
\bra{\lambda'}s_{\mathrm{ldn}}^{\beta}\ket{\lambda}=\begin{cases}
-\frac{1}{4}\bra{\lambda'}S_{\mathrm{ld}}^{\beta}\ket{\lambda}\quad &n=1\\
\frac{5}{12}\bra{\lambda'}S_{\mathrm{ld}}^{\beta}\ket{\lambda}\quad &n=2,3,4
\end{cases}
\end{align}
We could use the cluster spin operator $S_{ld}^{\alpha}$ and an effective form factor absorbing the above coefficient and the extra phase factors due to the displacement between coordinates of Cu-spin $\mathbf{r}_{ldn}$ and the "center of mass" coordinate $\mathbf{r}_{ld}$ representing the position of cluster. With the two approximations introduced above, the effective form factors of the spin clusters read
\begin{align}\label{Form}
\tilde{F}_{d}(\mathbf{Q})&=\left(-\frac{1}{4}e^{i\mathbf{Q}\cdot(\mathbf{r}_{\mathrm{ld1}}-\mathbf{r}_{\mathrm{ld}})}+\frac{5}{12}\sum_{i=2}^4e^{i\mathbf{Q}\cdot(\mathbf{r}_{\mathrm{ldi}}-\mathbf{r}_{\mathrm{ld}})}\right)F(\mathbf{Q})
\end{align}
The calculated cross-section in this cluster picture now reads
\begin{align}
& \frac{d^2\sigma}{d\Omega dE_f}(\mathbf{Q},\omega)=\frac{k_f}{k_i}(\gamma r_0)^2\sum_{\alpha,\beta}\left(\delta_{\alpha\beta}-\hat{Q}_{\alpha}\hat{Q}_{\beta}\right)\\&\times\sum_{\mathrm{l,d,l',d'}}\left(\tilde{F}_{d'}(\mathbf{Q}) e^{i\mathbf{Q}\cdot\mathbf{r}_{\mathrm{l'd'}}}\right)^*\left(\tilde{F}_{d}(\mathbf{Q})e^{i\mathbf{Q}\cdot\mathbf{r}_{\mathrm{ld}}}\right)\nonumber\\&\times\sum_{\lambda,\lambda'}p_{\lambda}\bra{\lambda}
S^{\alpha}_{\mathrm{l'd'}}\ket{\lambda'}\bra{\lambda'}S_{\mathrm{ld}}^{\beta}\ket{\lambda}\delta(E_{\lambda}-E_{\lambda'}+\hbar\omega)\nonumber
\end{align}
In Fig.~\ref{app2} (e,f,g) and Fig.~\ref{app2} (h,i,j), we compare constant energy slices  through the measured and calculated inelastic scattering cross section respectively for $\hbar\omega=3.0(3)$ meV. The excellent agreement validates the form factor we have derived.\par

We then carry out a pixel to pixel fit based on the form factor Eqn.~\ref{Form}. That is, we vary the parameters while respecting the $k_h$ constraints (0.0145(11) r.l.u. along $\langle 100\rangle$ directions\cite{adams2012long}) to minimize
\begin{align}\label{fitchi}
\chi^2=\frac{1}{N_{\text{pixels}}}\sum_i\frac{(Cy_i^{\text{cal}}-y_i^{\text{exp}})^2}{\sigma^2_i}
\end{align}
In the actual fit we loosened the constraint range for $k_h$ to $\pm~0.0033$ r.l.u considering the simplified nature of our model. Here $i$ labels the pixels in the experimental data (shorthand for $\mathbf{Q}$,$\hbar\omega$), $y_i^{\text{cal}},y_i^{\text{exp}},\sigma_i$ are calculated, measured cross sections and experimental errors, respectively. The constant of proportionality $C$ is determined by fitting $S(\mathbf{Q})$ as explained in the main text.\par

\section{Details of the quantitative comparison}\label{Fitting}
\subsection{Resolution Function and Broadening Factor}\label{reso}
A polynomial fit to the Monte Carlo simulated $\hbar\omega$-dependent energy resolution of the SEQUOIA instrument is shown in Fig~\ref{app2}(a). Energy cuts at three $R$ points with the 2-gaussian peak fit are shown in Fig~\ref{app2}(b,c,d). Since the magnon group velocity vanishes at this high symmetry point, momentum resolution contributions to the measured spectral line width vanish.  The FWHM of the lower peak at $\hbar\omega=8.4(1)$ meV is 0.51(9) meV, which exceeds the calculated instrumental resolution (of 0.34 meV).

Possible physical origins of the additional broadening are down-folding resulting from the incommensurate magnetic order, two magnon decay, magnon decay due to magneto-elastic interactions, magnon scattering associated with static or dynamic phase slips in the incommensurate order, and chemical inhomogeneity or disorder in the sample. While these mechanisms should generally be expected to be energy and momentum dependent, we treat them on average by adding a phenomenological relaxation rate in quadrature to the calculated energy resolution of the instrument: 
\begin{equation}\label{adjustresolution}
\tilde{\Delta}(\hbar\omega)=\sqrt{\Delta(\hbar\omega)^2+(2\bar{\Gamma})^2}.
\end{equation}
 Here $\Delta(\hbar\omega)$ is nominal FWHM energy resolution of the instrument and $2\bar{\Gamma}=0.37$ meV is the average phenomenological relaxation rate. $2\bar{\Gamma}$ is chosen so that  $\tilde{\Delta}(\hbar\omega)$ fits the FWHM of the lower peak at the $R$ point. $\tilde{\Delta}(\hbar\omega)$ is then used throughout the fitting analysis as the gaussian FWHM width of all modes.\par

\subsection{Reliability of Fitting Results}\label{Reliability}
Due to the limitations discussed in main text, the effective model can not describe all feaures in the measured neutron scattering  cross section. The set of parameters reported in the main text yields the global minimum of Eq.~\ref{fitchi}  $\chi^2_{min}\approx 13.26$. Here we evaluate the constraints that our data place on these parameters based on other sets of fit parameters yielding $\chi^2\leq\chi^2_{min}+5$. The upper limit corresponds to the analytical estimate in the main text ($J_1=-0.605$ meV, $J_2=-0.905$ meV, $d_1=-d_1'=0.2$ meV). The range for each DM component was chosen to be $[-0.6,0.6]$ meV, as these components must be significantly smaller than the corresponding Heisenberg exchange interactions.
\subsubsection{$J_1$ and $J_2$}
\begin{figure} [!htbp] 
	\includegraphics[width=0.85\columnwidth]{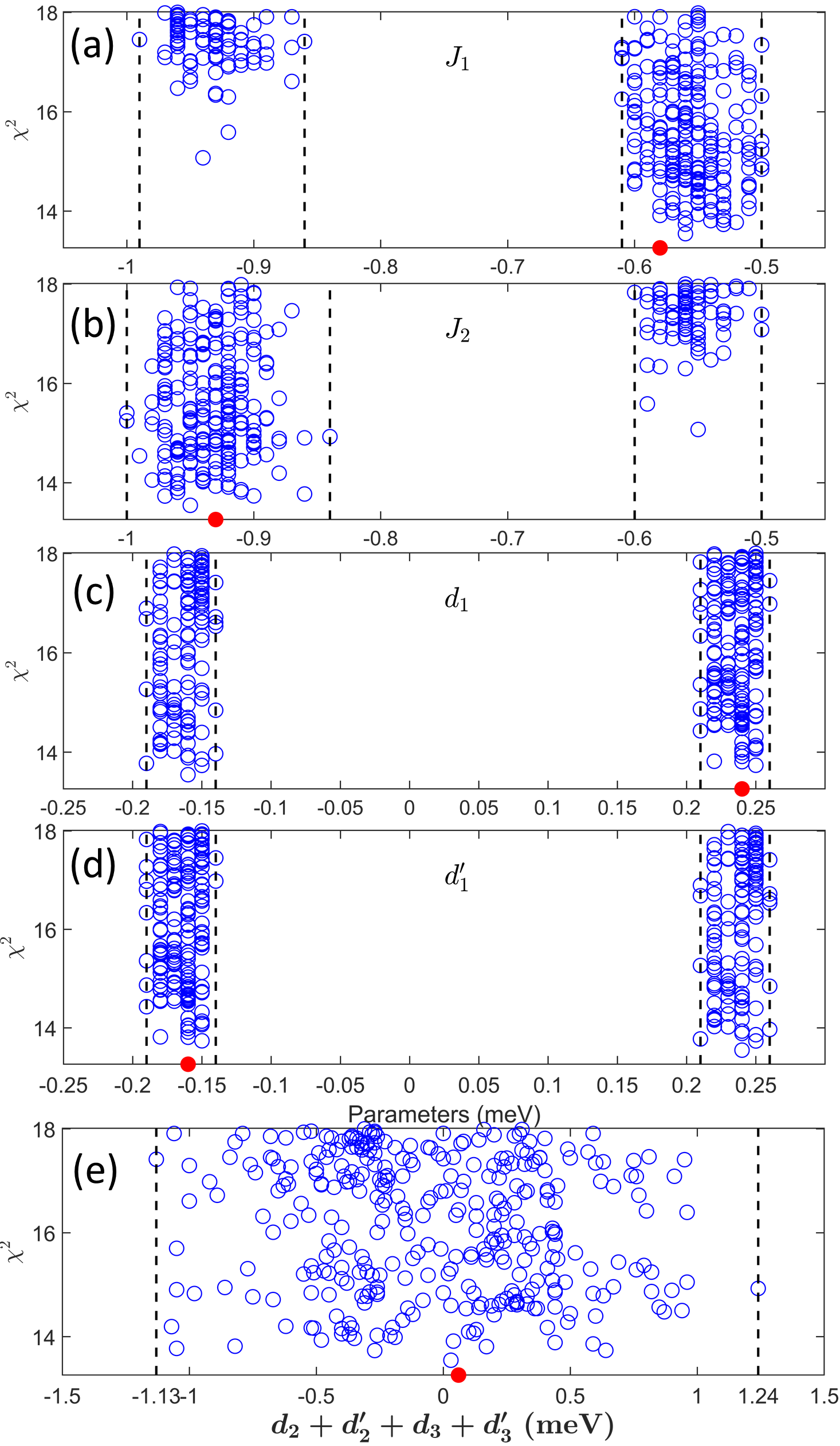}
	\caption{The projections of the goodness of fit $\chi^2$ on the parameter axis for (a) $J_1$, (b) $J_2$, (c) $d_1$, and (d) $d_1^{\prime}$. Each blue circle represents one set of parameters with low $\chi^2$. The red circle shows the optimal set of parameters listed in Table \ref{Pa}, the dashed lines show the bounds in these parameters inferred from the $\chi^2$ plots. (e) The projection of  $\chi^2$ on the parameter axis $d_2+d_3+d_2'+d_3'$.}
	\label{app3}
\end{figure}

As mentioned in the main text, the pixel to pixel fit must compromise between fitting the $\Gamma$ point and $R$ point, which leads to a range of $J_1$ and $J_2$ with comparable $\chi^2$. Also, the relative strength of $|J_1|$ and $|J_2|$ can not be determined, the fit provides the following bounds: $1.35 \leq|J_1+J_2|\leq 1.55$ meV and $0.3 \leq|J_1-J_2|\leq 0.5$ meV, which are related to the bandwidth of the magnon band at the $\Gamma$ point and the splitting at the M point, respectively, as described in Section \ref{theo}. The best fit is achieved when $|J_1|<|J_2|$ with experimental bounds on $J_1$ and $J_2$ as listed in Table \ref{Pa} and shown in Fig.~\ref{app3}(a,b).\par 
\subsubsection{$d_1$ and $d_1'$}
Fig.~\ref{app3}(c,d) shows that the DM components $d_1$ and $d_1^{\prime}$ lie in the range -0.19~meV$\leq d_1\leq -0.14$ meV and 0.21~meV$\leq d_1^{\prime}\leq 0.26$~meV, or interchangably 0.21~meV$\leq d_1\leq 0.26$ meV and -0.19~meV$\leq d_1^{\prime}\leq -0.14$ meV, with the rough constraint 0.04~meV$\leq d_1+d_1^{\prime}\leq 0.12$~meV. The ranges for $d_1$ and $d_1^{\prime}$ result from (1) the analytical relationship $|d_1-d_1'|\approx 0.4$ meV from Sec.~\ref{Rgap}. (2) the constraint from $k_h\propto (d_1+d_1^{\prime})$. $d_2,d_2^{\prime},d_3,d_3^{\prime}$ play secondary roles in determining $k_h$. The positive sign of $(d_1+d_1^{\prime})$ ensures a right-handed magnetic helicoid for a right-handed enantiomer and vice versa.\par
\subsubsection{$d_2,d_2^{\prime},d_3,d_3^{\prime}$}\label{alge}
\begin{figure*} [!htbp] 
	\includegraphics[width=0.9\textwidth]{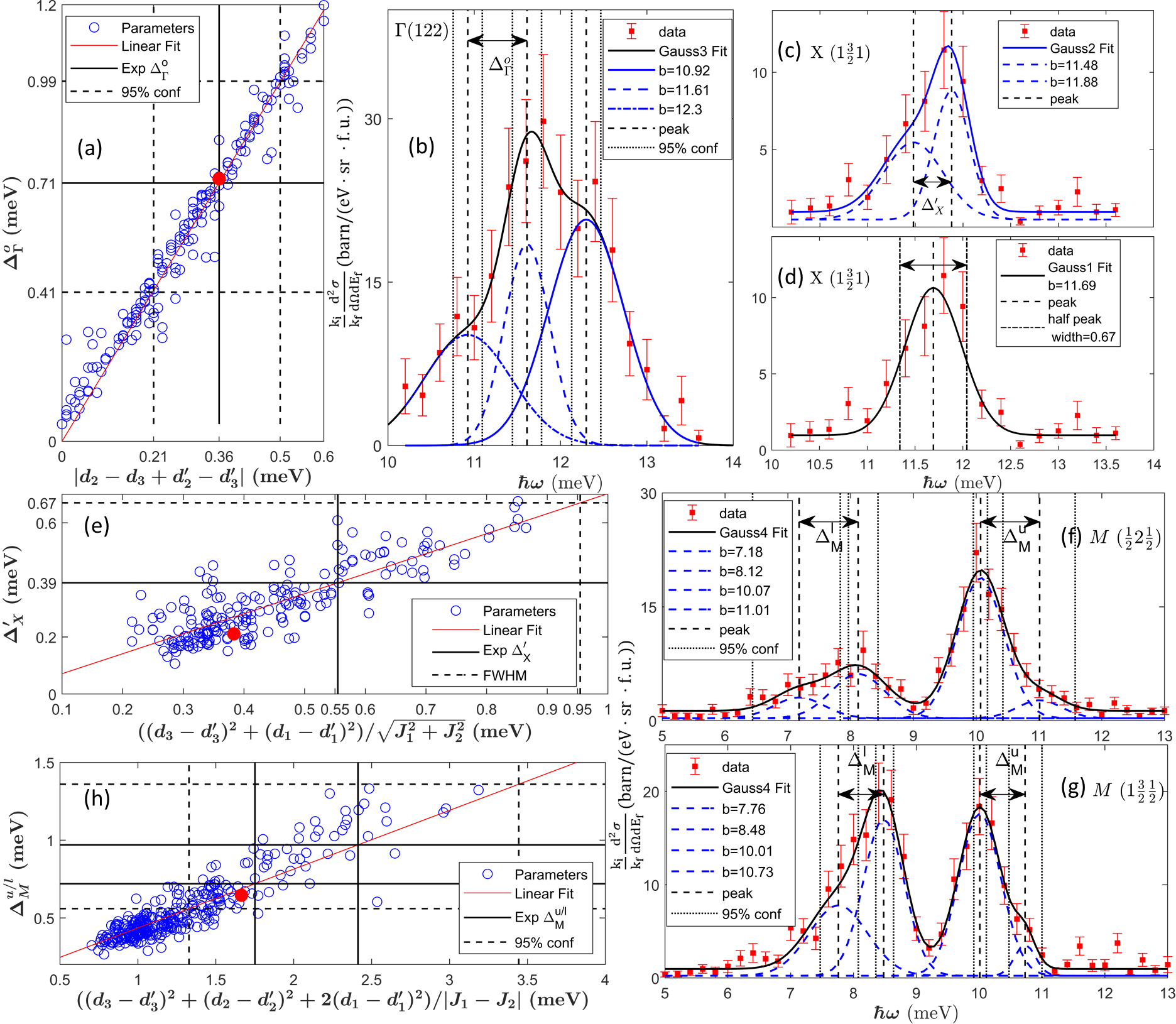}
	\caption{(a) The linear dependence of the optical mode splitting at the $\Gamma$ point $\Delta_{\Gamma}^{o}$ on $|d_2+d_2'-d_3-d_3'|$. The linear fit is $y=ax+b$ with $a=0.45(2)$, $b=-0.03(3)$ meV, with $R^2=0.85$. (b) Excitation spectrum at the $\Gamma$ point which provides experimental evidence for  $\Delta_{\Gamma}^{o}$ with 95\% confidence range shown. (c,d) 2-gaussian and 1-gaussian fits of the upper modes at the $X$ point around 12 meV. The splitting $\Delta_X^{\prime}$ is defined in (c). (e) The linear correlation of $\Delta_X^{\prime}$ and the quantity $D_X$ defined in Sec.~\ref{alge}(2). The linear fit is $y=ax+b$ with $a=0.70(6)$, $b=0.00(3)$ meV, with $R^2=0.71$. (f,g) 4-gaussian fit of two different $M$ point spectra, from which we obtain $\Delta_{M}^{l/u}=0.97(39)$ meV and $\Delta_{M}^{l/u}=0.7(2)$ meV, respectively. (h) The linear correlation of $\Delta_M^{u/l}$ and the quantity $D_M$ defined in Sec.~\ref{alge}(3). The linear fit is $y=ax+b$ with $a=0.45(2)$, $b=-0.03(3)$ meV, with $R^2=0.85$. In (a,e,h), we have marked the optimal set of parameters in Table \ref{Pa} by the red solid symbol. Error bars in all figures represent one standard deviation. In (b-d,f,g), red symbols show  neutron intensity data averaged over $(0.1~\text{\AA}^{-1})^3\times(0.2~\text{meV})$ in the 4D $\mathbf{Q}-\hbar\omega$ space.}
	\label{app4}
\end{figure*}
Our experiment establishes correlated constraints on $d_2,d_2',d_3,d_3'$ that relate to specific features in the data.\par

(1) $2|d_2+d_2'-d_3-d_3'|\approx \Delta_{\Gamma}^{o}$. This quantity corresponds to the splitting of optical modes at the $\Gamma$ point, as shown in Fig.~\ref{app4}(b). The optical modes are degenerate without DM interactions, and roughly speaking split into three modes with symmetric spacing $\Delta_{\Gamma}^{o}$ when DM interactions are turned on. The gaussian fits yield a mode splitting of 0.7(3) meV, which implies that  $|d_2+d_2'-d_3-d_3'|\approx 0.35(15)$ meV (Fig.~\ref{app4}(a)). As expected, $d_1,d_1'$ play no significant roles in this splitting.\par
\begin{figure*} [!htbp] 
	\includegraphics[width=0.85\textwidth]{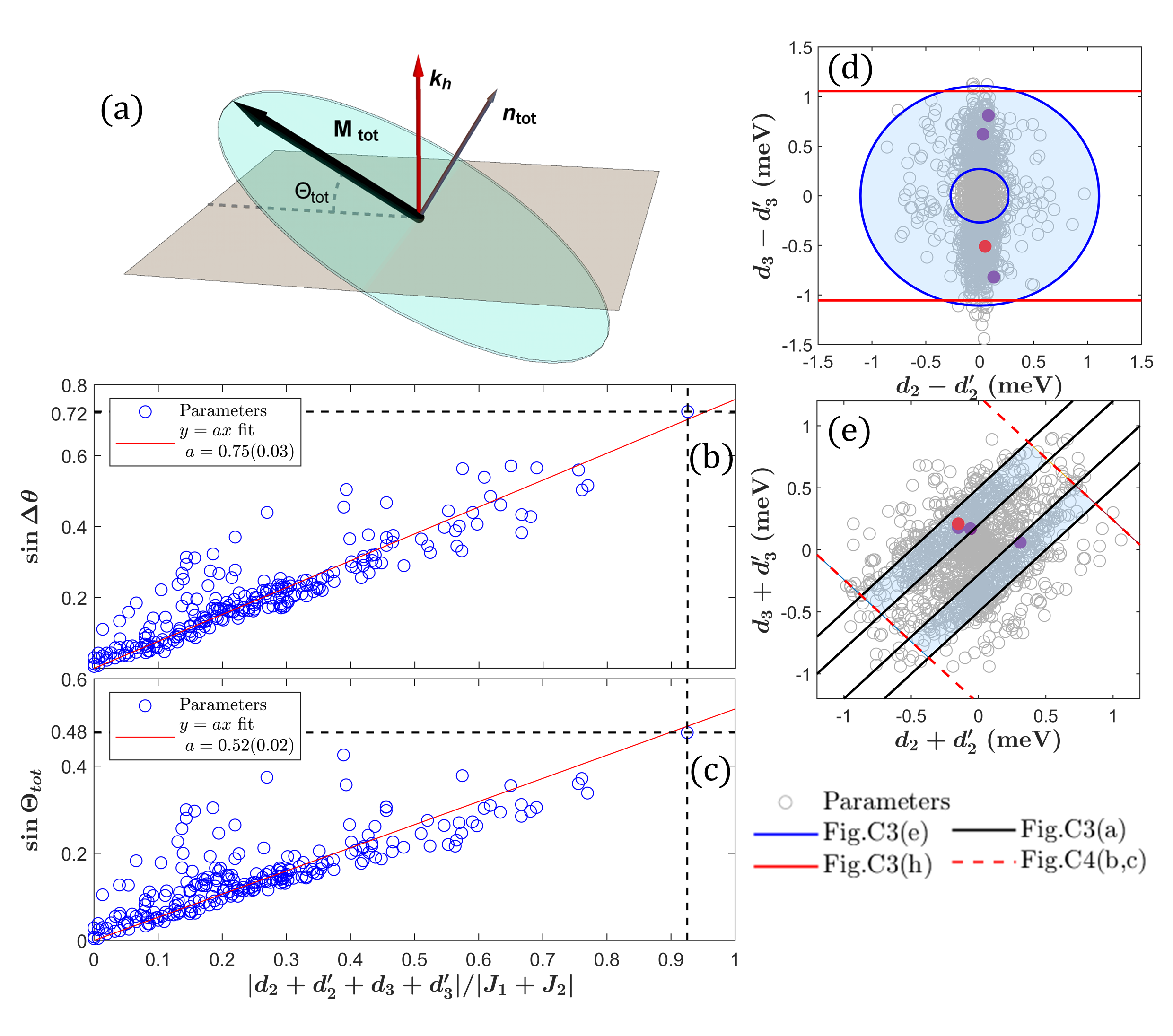}
	\caption{(a) "Tilting" helical state discussed in Sec.~\ref{alge}(4). $\mathbf{M}_{tot}$ represents the total magnetization of a unit cell (local magnetization density). The light green plane represents the precession plane of $\mathbf{M}_{tot}$, while $\mathbf{n}_{tot}$ is its normal direction. $\Theta_{tot}$ is the tilting angle of the precession plane with respect to the plane perpendicular to $\mathbf{k}_h$. (b,c) The linear correlation between (b) $\sin \Delta \theta$ and (c) $\sin \Theta_{\text{tot}}$ with respect to the dimensionless quantity $|d_2+d_2^{\prime}+d_3+d_3^{\prime}|/|J_1+J_2|$. Larger $|d_2+d_2^{\prime}+d_3+d_3^{\prime}|$ result in larger canting among spins on neighboring sublattices and and a larger tilt angle with respect to the transverse plane. (d) Graphic representation of constraints on $d_2,d_2',d_3,d_3'$ by dispersion analysis at the $M$ point (blue, Fig.~\ref{app4}(h)) and the $X$ point (red,Fig.~\ref{app4}(e)).(e) Graphic representation of constraints on $d_2,d_2',d_3,d_3'$ by analysis of the spectrum at the $\Gamma$ point (black,Fig.~\ref{app4}(a)). The red dashed line shows the constraint from (b,c). We assume the spin canting along $\mathbf{k}_h$ is small ($\Theta_{\text{tot}}\lesssim 30^{\circ}$, $\Delta\theta\lesssim 46^{\circ}$). In (d,e), the parameters satisfying the constraints of $J_1,J_2,d_1,d_1^{\prime}$ (Fig.~\ref{app3}(a-d)) are plotted. The 4 sets of parameters listed in Table.~\ref{Parameter_com} are plotted in red (the lowest $\chi^2$ value in Table~\ref{Pa}) and purple circles. The subspace of parameters allowed by constraints are filled in light blue color. }
	\label{apptilt}
\end{figure*}

(2) $\Delta_X^{\prime}\propto((d_1-d_1')^2+(d_3-d_3')^2)/\sqrt{J_1^2+J_2^2}\equiv D_X $. $\Delta_X^{\prime}$ is defined as the splitting/broadening of the upper modes at the $X$ point, which are two-fold degenerate without DM interactions. These lift the degeneracy due to the associated symmetry breaking and the superposition of contributions from the $X$ and $Z$ points from multiple domains of the incommensurate magnetic order. Strictly speaking, we should observe more than two modes at the $X$ point around 12 meV. If we nonetheless fit the broad maximum peak with two gaussian peaks (Fig.~\ref{app4}(c)), we obtain a rough estimate of  $\Delta_X^{\prime}\approx 0.39$ meV. Alternatively, if we fit with one broad gaussian peak as in Fig.~\ref{app4}(d), we obtain FWHM $\approx 0.67(19)$ meV. These fits give similar $\chi^2=1.2$, and the FWHM of the peaks are near $\Delta(\hbar\omega)$ and $\tilde{\Delta}(\hbar\omega)$ introduced in Sec.~\ref{reso}. We use the FWHM of the single gaussian fit as the upper bound on $\Delta_X^{\prime}$. We observe a linear correlation between the quantity $D_X$ and $\Delta_X^{\prime}$, as shown in Fig.~\ref{app4}(e), which gives us the constraint $((d_1-d_1')^2+(d_3-d_3')^2)/\sqrt{J_1^2+J_2^2}\leq 0.96$ meV. The denominator $\sqrt{J_1^2+J_2^2}$ is proportional to the energy difference between the calculated upper and lower modes at the $X$ point when DM interactions are absent.  \par

\begin{center}
	\begin {table*} 
	\centering
	\begin{tabular}{|| l c c c c c c c c c c||} 
		\hline
		\multirow{2}{*}{Parameter Sectors}& \multicolumn{8}{c}{Parameter(meV)}&\multicolumn{2}{c||}{Calculated Result}\\ [0.5ex]\cline{2-11} &$J_1$ & $J_2$ & $d_1$ & $d_2$ &$d_3$ &$d_1'$ &$d_2'$ &$d_3'$& $k_h$(r.l.u) & $\chi^2$\\ [0.5ex] 
		\hline\hline
				$|J_1|<|J_2|,d_1>0$&$-0.58^{+0.08}_{-0.03}$&$-0.93^{+0.03}_{-0.05}$&	$0.24^{+0.01}_{-0.03}$&	-0.05&	-0.15&	$-0.16^{+0.01}_{-0.03}$&	-0.10&	0.36&0.0143&13.26\\ 
		$|J_1|<|J_2|,d_1<0$&$-0.56_{-0.04}^{+0.06}$&$-0.95_{-0.05}^{+0.09}$&$-0.16_{-0.03}^{+0.02}$&	-0.06& 0.40&	$0.24_{-0.03}^{+0.02}$&	-0.09&	-0.22&0.0129&13.54\\ 
		$|J_1|>|J_2|,d_1>0$&$-0.96_{-0.03}^{+0.07}$&$-0.54_{-0.05}^{+0.03}$&$0.22_{-0.01}^{+0.04}$&-0.08&-0.36&$-0.18_{-0.01}^{+0.04}$&-0.14&0.42&0.0162&16.47\\ 
		$|J_1|<|J_2|,d_1<0$&$-0.94_{-0.02}^{+0.07}$&$-0.55_{-0.05}^{+0.03}$&$-0.15^{+0.01}_{-0.03}$&	0.22&-0.38&$0.25^{+0.01}_{-0.03}$&0.09&0.44&0.0151&15.04\\ 
			\hline\hline
		\cite{janson2014quantum} no spin-mixing &-0.65&-0.75&	0.09&	-0.08&	0.06&	-0.04&	-0.05&	0.00 &$\approx 0$&35.44\\ 
		\cite{janson2014quantum} spin-mixing &-1.09&-0.91&	0.14&	-0.14&	0.11&	-0.05&	-0.06&	0.00 &0.0011&143.09\\ 
		\cite{tucker2016spin} no spin-mixing&-0.86&-0.63&	0.09&	-0.08&	0.06&	-0.04&	-0.05&	0.00&0.0014&31.33\\
		\cite{tucker2016spin} spin-mixing&-0.86&-0.63&	0.14&	-0.14&	0.11&	-0.05&	-0.06&	0.00&0.0018&27.56\\ 
		\cite{zhang2019magnonic}&-0.65&-0.73&	0&	0&0&	0.10&	-0.08&	0.35&0.0196&29.71\\ 
		\hline
	\end{tabular}	
	\caption {Row 1-4: The optimal sets of parameters for 4 sectors and the range of confidence with all the constraints in Sec.~\ref{alge} applied. In this work we are unable to pin down the errorbars for each of $d_2,d_2^{\prime},d_3,d_3^{\prime}$. The constraints on these four parameters are discussed in Appendix.~\ref{alge}, the graphic representations are shown in Fig.~\ref{apptilt} (d,e). Row 5-9: Comparison of parameters from previous studies. The parameters are translated from references using Eqns.~\ref{translate}. For Ref.~\cite{janson2014quantum}, "spin-mixing"/"No spin-mixing" corresponds to two sets of parameters where spin-1/spin-2 admixture of cluster is considered (or not). The  $k_h$ and $\chi^2$ for rows 5-9 are calculated by our numerical method.
	 }\label{Parameter_com}
	\end {table*}
\end{center}

(3) $\Delta_M^{u/l}\propto(2(d_1-d_1')^2+(d_2-d_2')+(d_3-d_3')^2)/|J_1-J_2|\equiv D_M $. The two doublets at the $M$ point in Fig.~\ref{Fig2}(a)(magenta) are split into more than four modes due to the presence of multiple incommensurate magnetic domains. Furthermore, as previously discussed there is non-negligible broadening of the lower mode at the $M$ point that we ascribe to two-magnon decay processes.  The experimental limit on the splitting of the lower and upper doublets are denoted by $\Delta_M^{l}$ and $\Delta_M^{u}$, respectively. In the numerical calculation we find $\Delta_M^{l}\approx \Delta_M^{u}$. In Fig.~\ref{app4}(f,g), we fit two different $M$ points using two-gaussian models for each doublet. Due to the indefinite number of split modes for the incommensurate state, we loosen the constraint on the peak width to $1.4\tilde{\Delta}(\hbar\omega)$ so that the two-gaussian fit might accommodate multiple weaker split modes. The fit gives $\Delta_M^{l}\approx \Delta_M^{u}\in[0.52,1.36]$ meV. We observe a linear correlation between the quantity $D_M$ and $\Delta_M^{u/l}$, as shown in Fig.~\ref{app4}(h), which yields the constraint 1.31~meV$\leq(2(d_1-d_1')^2+(d_2-d_2')+(d_3-d_3')^2)/|J_1-J_2|\leq 3.09$~meV. The denominator $|J_1-J_2|$ is proportional to the energy difference between the upper and lower doublets at the $M$ point when DM interactions are absent.\par

(4) $|d_2+d_3+d_2'+d_3'|\leq 1.24$ meV, as shown in Fig.~\ref{app3}(e). This quantity is related to the  tilting of spins towards the direction of $\mathbf{k}_h$, which is different on each of the four sublattices. This quantity also appears in the field theory description of Ref.~\cite{janson2014quantum} ($\kappa$ term in Eqn. (5,6)). A large $|d_2+d_3+d_2'+d_3'|$ will give us a "tilting"  zero field helical state, with the magnetization precessing in a plane that is not perpendicular to $\mathbf{k}_h$. The non-uniform tilting will also result in a magnetic structure far from collinear even at the atomic scale, and yields a larger bandwidth of magnon dispersion than $8|J_2+J_2|$ predicted in Sec.~\ref{theo}. The linear correlation of spin canting between sublattices and tilting angle with the quantity  $|d_2+d_3+d_2'+d_3'|/|J_1+J_2|$ is shown in Fig.~\ref{apptilt}(b,c). For this work we assume that the spin canting along $\mathbf{k}_h$ is small in the zero field magnetic structure, the tilting angle $\Theta_{\text{tot}}\lesssim 30^{\circ}$ (see Fig.~\ref{apptilt}(a)), and that the local canting angles between neighboring spins $\Delta\theta\lesssim 46^{\circ}$. In this regime the bandwidth $\approx 8|J_2+J_2|$ and the correction of $|d_2+d_3+d_2'+d_3'|$ to the bandwidth is negligible. A polarized neutron diffraction experiment in a single domain state should be able to establish the degree of non-coplanarity without the need to actually resolve the incommensurate wave vector. \par

\subsubsection{Comparison to previous study}			
In Table.~\ref{Parameter_com} we compare our fit parameters to previous studies\cite{janson2014quantum,portnichenko2016magnon,tucker2016spin,zhang2019magnonic}. The microscopic parameters $J_w^{\text{AF}},J_w^{\text{FM}},J_{\text{o.o}}^{\text{AF}}$ and the DM interaction on these bonds can be transformed into FM exchange and DM interaction in the effective spin-1 cluster picture under the assumption $|J_s^{\text{AF}},J_s^{\text{FM}}\rightarrow\infty|$. The transformations (worked out in Ref.~\cite{janson2014quantum}) are 
\begin{align}
J_1&=-l_1l_2\left(J_w^{\text{AF}}+J_{\text{o.o}}^{\text{AF}}\right)\\
J_2&=l_2^2J^{\text{FM}}_w\nonumber\\
\left(d_1,d_2,d_3\right)&=-l_1l_2\left[\left(D^y_{\rho_1,\rho_8},D^z_{\rho_1,\rho_8},D^x_{\rho_1,\rho_8}\right)\right.\nonumber\\&\left.+\left(D^y_{\rho_4,\rho_{12}},-D^z_{\rho_4,\rho_{12}},D^x_{\rho_4,\rho_{12}}\right)\right]\nonumber\\
\left(d_1',d_2',d_3'\right)&=l_2^2\left(D^z_{\rho_5,\rho_{12}},D^x_{\rho_5,\rho_{12}},D^y_{\rho_5,\rho_{12}}\right)\nonumber\\
l_1&=\frac{1}{4}\quad l_2=\frac{5}{12}\nonumber
\end{align}\label{translate}
Notice in our spin-cluster picture we only consider finite $J_w^{\text{AF}},J_w^{\text{FM}},J_{\text{o.o}}^{\text{AF}}$. Reference \cite{janson2014quantum} (without spin-mixing) and references \cite{ozerov2014establishing} and \cite{portnichenko2016magnon} essentially give the same set of parameters. Reference \cite{tucker2016spin} gives a different set of exchange parameters $J_w^{\text{AF}},J_w^{\text{FM}},J_{\text{o.o}}^{\text{AF}}$ but it does not present new information about  DM interactions. In our comparison to these parameters, we use the same DM parameters as in reference \cite{janson2014quantum}. In Table~\ref{Parameter_com}, we include the optimal parameters for the 4 sectors of low $\chi^2$ fits distinguished by: (1) the relative strength of $|J_1|$ and $|J_2|$ and (2) the sign of $d_1$ and $d_1'$ (which should be opposite to each other), along with the error bars for each sector.\par

\begin{figure} 
	\includegraphics[width=3 in]{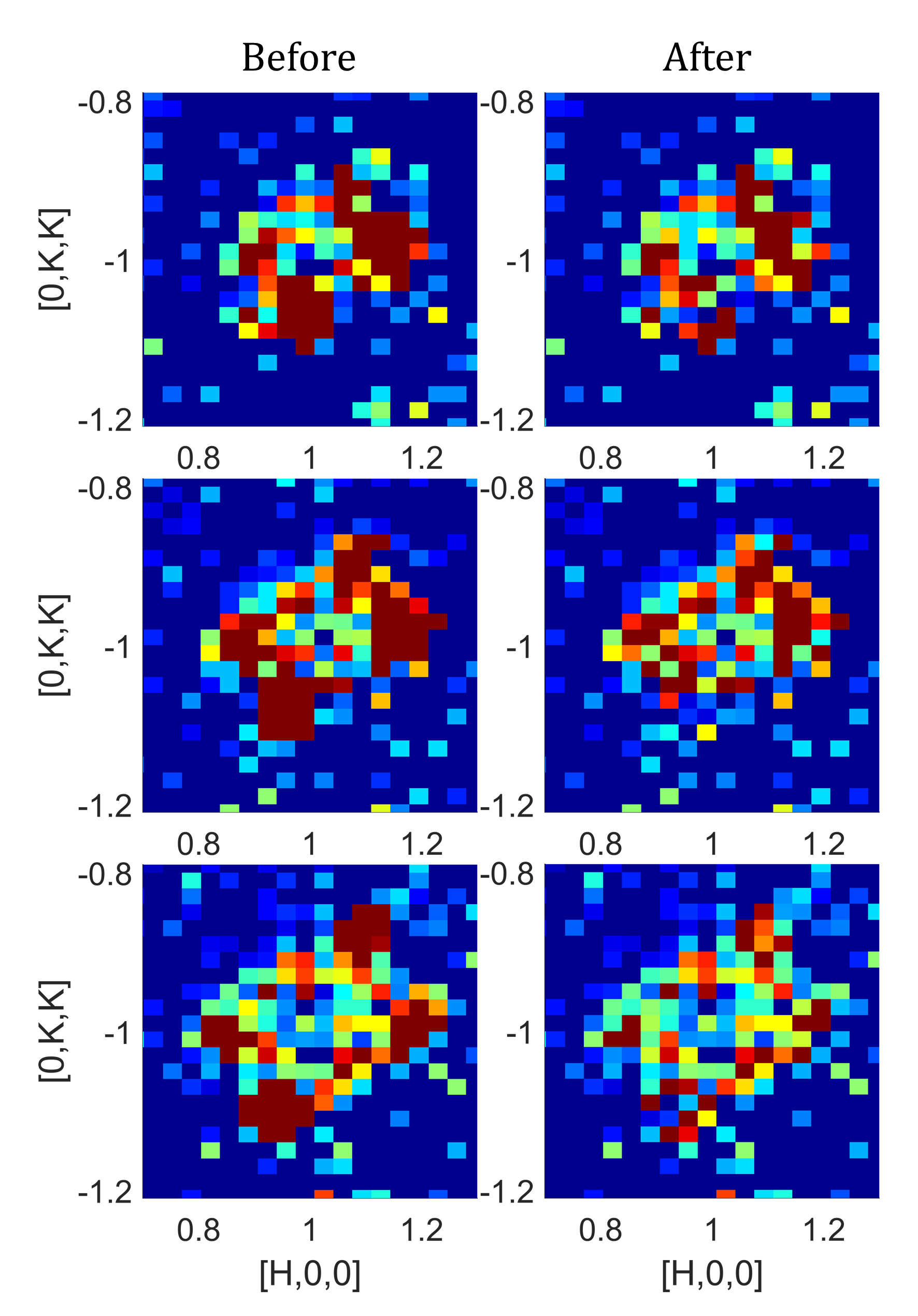}
	\caption{(a-c) Constant energy slices of MACS data at energy transfers $\hbar\omega=$ (a) 0.40(5), (b) 0.50(5), (c) 0.60(5) meV before the "spurion" subtraction, respectively. (d-f) Constant energy slices at energy transfers  $\hbar\omega=$ (d) 0.40(5), (e) 0.50(5), (f) 0.60(5) meV  after removing the Bragg "spurions", respectively.}
	\label{app6}
\end{figure}

\section{Details of MACS data analysis}\label{MACSapp}
\subsection{Subtraction of Bragg spurions}\label{spur}
During the processing of MACS data, we identified and subtracted Bragg spurions that arise when neutrons at the energy $E_f=2.4$ meV reach the sample due to a diffuse process at the monochromator and Bragg diffract from the sample. Such processes are more prominent on MACS than on conventional triple axis spectrometers because of the large monochromator and the lack of collimation between the monochromator and the sample. Bragg spurions occur in groups of four in symmetrized data because the spurions do not respect the mirror planes. In Fig.~\ref{app6}, we show several constant energy slices through MACS data before and after subtraction of the spurions.

\begin{figure*} 
	\includegraphics[width=6 in]{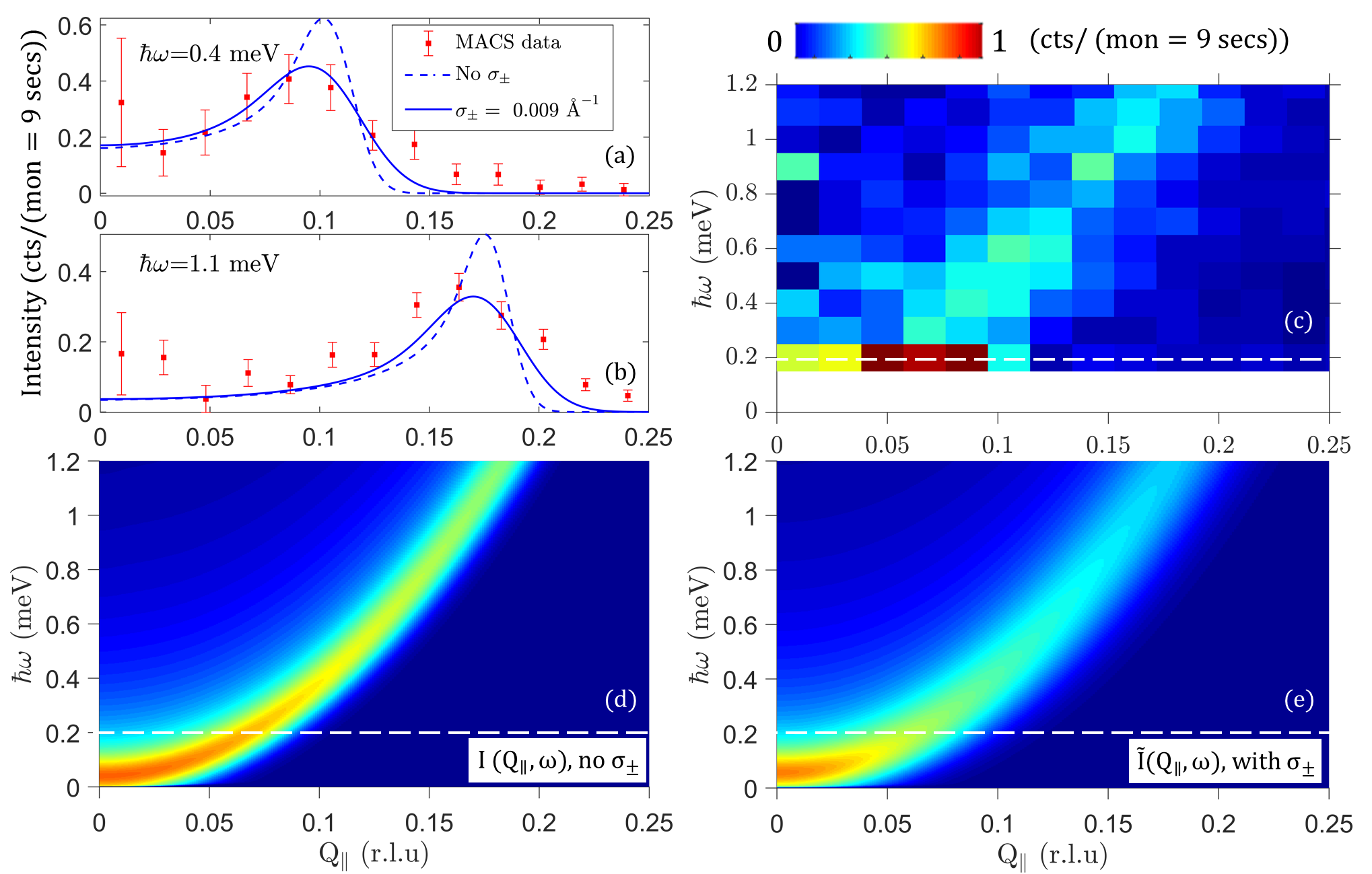}
	\caption{(a,b) Constant energy cut of MACS data and the best fitted $\tilde{I}(Q_{\parallel},Q_{\bot}=0,\omega)$ ($D=67(8)~\text{meV}~\mathrm{\AA^2}$ and $\Delta_{\Gamma}=0.0(1)~\text{meV}$) at $\hbar\omega=0.4,1.1~\text{meV}$, respectively. $\tilde{I}(\mathbf{Q},\omega)$ has taken into account the average out-of-plane Q-resolution $\sigma_{Q_{\bot}}=0.07~\mathrm{\AA}^{-1}$, the energy resolution $\sigma_{\omega}=0.05~\text{meV}$, the in-plane Q-resolution $\sigma_{Q_{\parallel}}=0.004~\mathrm{\AA}^{-1}$. Both $I(\mathbf{Q},\omega)$ (Eqn.~\ref{qreso}) and $\tilde{I}(\mathbf{Q},\omega)$ (Eqn.~\ref{qfinal})  with the extra broadening by $\sigma_{\pm}$ (representing the presence of incommensurate modes) are plotted. The relatively better agreement of the $\tilde{I}(\mathbf{Q},\omega)$ (solid line) with the MACS data shows that we have observed the incommensurate nature of spinwave modes. (c) $Q_\parallel-\omega$ intensity map of MACS data following azimuthal averaging around  ${\bf Q}_0$ (the same as Fig.~\ref{MACS1}(c)). (d,e) Simulated $I(Q_{\parallel},\omega)$  and $\tilde{I}(Q_{\parallel},\omega)$  with the parameters and resolutions specified as in (a,b). Dashed lines in (c,d,e) marks the lowest accessible energy transfer (0.2 meV) in the MACS experiment. Error bars in (a,b) represent one standard deviation.}
	\label{app7}
\end{figure*}

\subsection{Resolution and incommensurability on MACS}\label{MRcal}
For low energy inelastic scattering we used the MACS instrument at the NCNR with the monochromator in the sagittal focusing mode (vertical focusing only) and a fixed final energy of $E_f=2.4$~meV. The vertical divergence of the incident (scattered) beam was controlled by a 160 mm vertical slit  before the monochromator (the analyzer dimensions) to be 4 degrees (8 degrees) FWHM, which corresponds to a gaussian standard deviation $\sigma_{\bot}=0.07~\text{\AA}^{-1}$ for momentum transfer perpendicular to the scattering plane. The horizontal beam divergence was controlled by a 60 mm horizontal slit before the monochromator and by a 90' collimator after the sample. Combined with the 2 degree effective sample mosaic this lead to an approximately isotropic in-plane momentum resolution with $\sigma_{\parallel}=0.004~\mathrm{\AA^{-1}}$. The finite energy resolution  $\sigma_{\omega}=0.051~\text{meV}$ is approximated as uncorrelated with momentum resolution. The four dimensional gaussian resolution function is thus described by a diagonal resolution matrix with identical in-plane matrix elements\cite{Chesser_1973}.

We shall discuss the resolution effects associated with a resonant dispersive dynamic structure factor of the form ${\cal S}({\bf Q},\omega)={\cal S}({\bf Q})\delta(\hbar\omega-\epsilon(\mathbf{Q}))$, which depends only on the in-plane $Q_\parallel=|({\bf Q}-{\bf Q}_0)_\parallel|$ and out of plane $Q_\bot=|({\bf Q}-{\bf Q}_0)_\bot|$ distance from ${\bf Q}_0=(1\bar{1}\bar{1})$. Such data can be subjected to azimuthal averaging about ${\bf Q}_0$ and plotted versus $Q_\parallel$ as in Fig.~\ref{MACS1}. The corresponding resolution smeared intensity distribution in the $Q_\bot=0$  plane can be written as follows

\begin{widetext}
\begin{align}
I(Q_{\parallel},\omega)=\int \frac{Q_{\parallel}'dQ_{\parallel}' }{\sigma_{\parallel}^2}I_0\left(\frac{Q_{\parallel}Q_{\parallel}'}{\sigma_{\parallel}^2}\right )\exp\left(-\frac{Q_{\parallel}^2+Q_{\parallel}'^2}{2\sigma_{\parallel}^2}\right )\int \frac{dQ_{\bot}'}{2\pi\sigma_{\bot}\sigma_{\omega}}\exp\left(-\frac{Q_{\bot}'^2}{2\sigma^2_{\bot}}\right){\cal S}(Q_{\bot}',Q'_{\parallel})\exp\left(-\frac{\left(\epsilon(Q_{\bot}',Q'_{\parallel})-\hbar\omega\right)^2}{2\sigma^2_{\omega}}\right)\label{qreso}
\end{align}
\end{widetext}
Here $I_0$ is  the zeroth modified Bessel function of the first kind. For ferrimagnetic \cu2 we use $\epsilon(Q_{\bot}',Q'_{\parallel})=\Delta_{\Gamma}+D(Q_\parallel^2+Q_\bot^2)$ and ${\cal S}(Q_{\bot}',Q'_{\parallel})={\cal S}$.  The fit yields $D=67(8)~\text{meV}~\mathrm{\AA^2}$ and $\Delta_{\Gamma}=0.0(1)~\text{meV}$, which is consistent with the values of $D=58(2)~\text{meV}~\mathrm{\AA^2}$, $\Delta_{\Gamma}=0.00(5)~\text{meV}$ associated with the  parameters in Table~\ref{Pa}.  Fig.~\ref{app7}(a,b) shows  constant energy cuts of MACS data with the best fit
$I(Q_{\parallel},\omega)$ as a dashed line. There is clear evidence for physical broadening beyond the resolution of the instrument. 

To represent the incommensurate modes $\mathbf{q}\pm N\mathbf{k}_h$ (see Sec.~\ref{Dis}), we include a gaussian convolution along the radial direction, and take the spacing between $\mathbf{q}\pm \mathbf{k}_h$ mode ($\approx 0.0145\times 2$ rlu) as FWHM, that is, $\sigma_{\pm}=0.009~\mathrm{\AA^{-1}}$. The simulated in-plane intensity with this broadening factor included is 
\begin{align}
&\tilde{I}(Q_{\parallel},\omega)=\int \frac{dQ''_{\parallel}}{\sqrt{2\pi\sigma_{\pm}^2}}\exp\left(-\frac{(Q_{\parallel}-Q_{\parallel}'')^2}{2\sigma_{\pm}^2}\right) I(Q_{\parallel}'',\omega)\label{qfinal}
\end{align} 
An excellent fit is now achieved as shown by the solid lines in Fig.~\ref{app7} and as a color image in Fig.~\ref{MACS1}(d). While a higher resolution experiment is needed to resolve the details, the present data already shows signs of additional low $Q$  structure in the inelastic scattering as anticipated for an incommensurate state.

\bibliographystyle{apsrev4-1}
\bibliography{ref_draft}

\end{document}